\documentclass[11pt]{article}
\usepackage{t1enc}
\usepackage[left=2.5cm,right=2.5cm,top=2.5cm,bottom=2.5cm,includeheadfoot]{geometry}
\usepackage[french,english]{babel}
\usepackage{amsmath, amssymb,amsfonts,amsthm}
\usepackage[colorlinks=true, citecolor  = red, urlcolor = blue]{hyperref}
\usepackage{fancyhdr}
\usepackage{fancybox}
\usepackage{xcolor}
\usepackage{float}
\usepackage{booktabs}
\usepackage{xspace}
\usepackage{todonotes}
\usepackage{graphicx}
\usepackage{enumerate}
\usepackage{eurosym}
\usepackage{amsthm}
\usepackage{todonotes}
\usepackage{bbm}
\usepackage[ruled,vlined]{algorithm2e}
\usepackage{mathtools}     
\usepackage{multirow}
\mathtoolsset{showonlyrefs=true}                                    
\newcommand{\otoprule}{\midrule[\heavyrulewidth]}

\usepackage{xparse}
\let\realItem\item 
\makeatletter
\NewDocumentCommand\myItem{ o }{%
   \IfNoValueTF{#1}%
      {\realItem}
      {\realItem[#1]\def\@currentlabel{#1}}
}
\makeatother

\usepackage{enumitem}    
\setlist[enumerate]{
    before=\let\item\myItem,       
    label=\textnormal{(\arabic*)}, 
    widest=(2')                    
}

\newtheorem{theorem}{Theorem}[section]
\newtheorem{proposition}[theorem]{Proposition}

\newtheorem{lemma}[theorem]{Lemma}

\newtheorem{definition}[theorem]{Definition}

\numberwithin{equation}{section}

\newtheorem{example}[theorem]{Example}

\newcommand \remark[1]{\begin{description}\item[Remark.]#1\end{description}}

\usepackage{framed}

\allowdisplaybreaks 

\newcommand{\dd}{{\rm d}} 
\newcommand \du{{\rm d}u}

\newcommand \BS{{\tt BS}} 
\newcommand \IV{{\tt IV}} 
\newcommand \bIV{\underline{\tt IV}} 
\newcommand \Der{\tt Der.} 
\newcommand\CallDer{{\tt Call}^{\Der}} 
\newcommand\Call{{\tt Call}}

\newcommand \C{{\tt C}} 
\newcommand \F{{\tt F}} 
\newcommand\Fut{{\tt Fut}} 
\newcommand\FutDer{\Fut^{\Der}} 
\newcommand\FutDerT{\Fut^{{\Der},T}} 
\newcommand\USD{{\tt USD}}
\newcommand\BTC{{\tt BTC}}
\newcommand\ETH{{\tt ETH}}
\newcommand\Size{{\tt S}}
\newcommand \bps{{\tt bps}}
\newcommand \vega{{\tt Vega}}
\newcommand \vomma{{\tt Vomma}}
\newcommand \Bid{\mathrm{Bid}}
\newcommand \Ask{\mathrm{Ask}}
\newcommand \Eff{\mathrm{Eff}}
\newcommand{\md}{\textup{\texttt{md}}}
\newcommand{\spr}{\textup{\texttt{sp}}}
\newcommand \anc{\mathrm{Anc}}
\newcommand \ts{\textup{\texttt{ts}}}
\newcommand \pie{\pi^{aug.}}

\newcommand {\wz} {w(\tau,k)}
\newcommand \rw{w_\tau(k;\chi^R)}

\newcommand{\N}{\mathbb{N}\xspace} 
\newcommand{\R}{\mathbb{R}\xspace} 
\renewcommand{\P}{\mathbb{P}} 
\newcommand \cN{\mathcal{N}}
\newcommand \1{\mathbf{1}}
\renewcommand\epsilon{\varepsilon}
\newcommand{\isdef}{\stackrel{\mbox{\tiny def}}{=}}

\newcommand\Esp[1]{\mathbb{E}\left[#1\right]}
\newcommand\Var[1]{\mathbb{V}{\rm ar}\left[#1\right]}
\newcommand{\var}{\mathbb{V}{\rm ar}} 

\newcommand \BF{B^\F}
\newcommand \BC{B^\C}
\newcommand \CF{C^\F}
\newcommand \CC{C^\C}

\allowdisplaybreaks

\newcommand{\setof}[2]{\left\{#1\,\middle|\:#2\right\}}

\title{Unbiasing and robustifying implied volatility calibration\\ in a cryptocurrency market\\ with large bid-ask spreads and missing quotes}

\author{Mnacho ECHENIM\footnote{Laboratoire d'Informatique de Grenoble (LIG), CNRS, Grenoble INP, UGA. 700 avenue Centrale, Domaine Universitaire, 38401 Saint Martin d'Hères, France. Email: \textsc{mnacho.echenim@univ-grenoble-alpes.fr}} \and Emmanuel GOBET\footnote{Centre de Math\'ematiques Appliqu\'ees (CMAP), CNRS, Ecole Polytechnique, Institut Polytechnique de Paris. Route de Saclay, 91128 Palaiseau Cedex, France.  Email: {\sc emmanuel.gobet@polytechnique.edu}} \and Anne-Claire MAURICE\footnote{Kaiko - Quantitative Data. 2 rue de Choiseul 75002 Paris, France. Email: \textsc{anne-claire.maurice@kaiko.com}}}
\date{\today}

\begin{document}

\maketitle

\begin{abstract}
  We design a novel calibration procedure that is designed to handle the specific characteristics of options on cryptocurrency markets, namely large bid-ask spreads and the possibility of missing or incoherent prices in the considered data sets. We show that this calibration procedure is significantly more robust and accurate than the standard one based on trade and mid-prices.   
    
    \smallskip	\noindent {\bf Keywords:} implied volatility, calibration, bid-ask spread, missing data, data augmentation

    \smallskip	\noindent {\bf MSC2020:} 91G20, 62G35,
    62G09, 62D10, 62P05

\smallskip	\noindent {\bf JEL:} G13, C80

\end{abstract}

\section{Introduction}
\paragraph{Context.} European option prices are traditionally  quoted in terms of the corresponding implied volatility, in order to have an adimensional measure of how expensive the option is, independently of the price magnitude of the underlying asset. Over the past two decades, a large number of papers (both from practitioners and academics) have focused on:
\begin{itemize} \item deriving some closed-form approximations of this price when the underlying asset is modeled using a specific  volatility model (local, stochastic, local-stochastic, with jumps, etc\ldots), see, e.g., \cite{haga:kuma:lesn:wood:02}, \cite{lee:05}, \cite{alos:leon:vive:07},  \cite{fouq:papa:sirc:knut:11}, \cite{berg:guyo:11}, \cite{gath:jacq:11}, \cite{bomp:gobe:12a}, \cite{lori:pagl:pasc:17} among many others. For an overview, see the book of Gatheral \cite{gatheral:volsurface:11}; other references include \cite{bomp:gobe:12b} or \cite{friz:gath:guli:jacq:teic:15};
\item understanding how to calibrate simplified parametric forms (such as SVI \cite{gath:04} or its extensions \cite{corb:coho:laac:mart:19}) with the goal of designing robust optimization techniques, with good initialization guesses \cite{dema:mart:09,lefl:14a,corb:coho:laac:mart:19}.
  \end{itemize}
This is now a mature research area when applied to usual markets like equity/index, fixed income, fiat foreign exchange, or commodities. The aim of this work is to adapt these techniques to cryptocurrency markets, which have grown at a huge pace over the past years.

\paragraph{Crypto options history.}
Cryptocurrency markets have gone through an unprecedented period of growth, and the demand for sophisticated trading products has surged. Previously, options trading was limited to traditional financial markets, but there is an increasing adoption of options throughout cryptocurrency markets. 
Options trading for retail investors seems to align well with the speculative nature of retail crypto trading and as a speculative trading product, crypto options have grown at a similar pace as the broader crypto market. 
Speculation is not the only use case however, particularly for institutions, as options contracts also offer the ability to hedge against the risk of sudden price changes in crypto markets.

The exchange with the most market share of options trading is Deribit; the company claims an 80 to 90\% market share of all options trading. Deribit was launched in 2016 and as a result certainly benefited from first-mover advantage in the market. Deribit option trading products have continued to become more and more complex, moving in line with the increased liquidity in futures markets as well as the increased demand for complexity in crypto trading products. This added complexity and continued stream of new products has allowed Deribit to maintain their market share, fending off competition from other exchanges.

The Chicago Mercantile Exchange launched regulated crypto options in 2020, and this marked an important milestone in the industry, paving the way for institutional adoption of crypto options. In response, Deribit launched daily traded options shortly after in order to satisfy the increased market demand for more complexity. 2022 was the first time an over-the-counter options trade was made by a major U.S bank, as Goldman Sachs executed a trade involving a non-deliverable (cash settled) bitcoin option. This transaction was evidence of a growing institutional presence and gives an insight into how the biggest players in traditional finance are increasingly utilizing crypto options today.

\begin{figure}[t]
\begin{center}\includegraphics[scale=0.5]{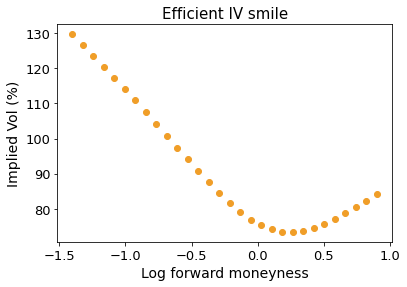}
\caption{Implied volatility in the Raw SVI model \eqref{eq:raw:par} with the parameters $(a,b,\rho, m,\sigma)=(0.02, 0.2, -0.3, 0.05,0.6)$.}\label{fig:raw:svi}
\end{center}
\end{figure}

\paragraph{On calibration.} 
Traditionally, fitting a smile volatility {(with time-to-expiry $\tau$)} model boils down to finding the optimal parameter $\chi$
 that minimizes the Mean Square Error (MSE)
\begin{align}\sum_{i}\left( {w(\tau,k_i;\chi)}- w^{\tt Mkt }(\tau,k_i)\right)^2,
\label{eq:MSE}
\end{align}
where:
\begin{itemize} 
\item each $k_i$ is a log-forward-moneyness at which market data {for time-to-expiry $\tau$} is available;
\item $k\mapsto w(\tau,k;\chi)$ is a parametrization of the total variance (the squared implied volatility multiplied by the time-to-expiry), such as, e.g.,  the arbitrage-free SVI parametrization presented in \eqref{eq:raw:par} and depicted in Figure \ref{fig:raw:svi} for a given set of parameters;
\item $w^{\tt Mkt }(\tau,k_i)$ is the market total variance at $k_i$.
\end{itemize}

The MSE is not the only loss criterion that practitioners tend to consider:  the $L_2$-norm can be replaced by the  Huber loss \cite{hube:64}, or simply the $L_1$ norm; the errors may be evaluated on the implied volatilities or the prices instead of total variances;  weighted sums can also be considered, with weights that depend on the volume. 

\paragraph{Statement of the problem.}
Our paper sheds light on an issue that is often ignored on traditional markets (where it is negligible), but has a significant impact in crypto markets. Usually the quantity $w^{\tt Mkt }(\tau,k_i)$ is computed based on trade and mid-prices, the latter being an excellent approximation of the so-called \textit{efficient price} when the bid-ask spread is small. However, it appears that the relative tick size  (and thus the relative spread) for options on cryptos is roughly 10 to 50 times larger than for options on  traditional markets (see Table \ref{table:compare:ticksize}). Hence the mid-price appears to be a very noisy estimate of the efficient price, and as a consequence it may result in a higher error of the calibrated implied volatility. Besides, it is not uncommon for  quotes to be incomplete (only the ask or the bid is available), and this differs with usual markets where assigned market-makers play the role of liquidity providers.

On the other hand, instead of using the mid-price information, one could restrict the considered data to trades. Unfortunately, at the time of writing this article, the option market liquidity  is limited, which is why it is necessary to use the bid-ask information in the calibration scheme in order to have more stable results.
As an illustration, over one week (from June 27 to July 3, 2022), we counted the number of BTC-USD option trades on the Deribit platform: there was an average of 30 trades per hour, all strikes combined, for the most liquid expiries (1 week and 1 month), around 5 trades per hour for the quarterly 3-month and 6-month expiries and that only 1 or 2 trades for the 9-month and 1-year expiries.
The need for all trade and bid-ask prices is also reinforced because the time scale of the crypto market is shorter than that of traditional markets: indeed this is a market that is open 24 hours a day, 7 days a week. In addition, the volatility (resp. total variance) is around three (resp. nine) times higher than on equity/index markets, highlighting the need to update market prices more often, typically every 2 to 3 hours instead of once a day. 
The data that is used throughout the paper was provided by Kaiko\footnote{\url{https://www.kaiko.com/}}, whose clients have shown a strong interest in obtaining calibrated implied volatilities at an hourly scale.

\paragraph{Our contributions.}
For such situations where tick sizes are relatively large (and thus spreads are also large), one may try to use the bid and ask prices, instead of only the mid-price. In a nutshell, how to adapt the calibration criterion written for mid-prices to account for bid and ask prices that may differ a lot  from this mid-price? How to handle cases where one side of the quote is missing, discarding any attempt to use the naive mid-price? To the best of our knowledge, these questions are overlooked in the literature, which can be easily understandable in view of the usually small spreads and tick sizes of standard markets. 
Our contributions are the following.
\begin{itemize} \item  The market implied volatility is much more sensitive to large spreads when computed from ITM puts or from ITM calls, see Figure \ref{fig:sensitivity:IV} for a precise statement. As a result, we filter call and put prices taking only OTM options.

\item We prove that calibrating on the mid-price yields a systematic  underestimation of the implied volatility parametrization in the wings (for small or large log-forward-moneynesses), see Theorem \ref{th:biais} for a precise statement.

\item In order  to   remove  the  bias in the calibration process, we define a so-called \emph{anchor price}, which is built from the bid and ask prices, and has a negligible bias (of order three w.r.t. the tick size) with respect to the efficient price. See Theorem \ref{th:debiais} for a precise statement.

\item In the case where there are missing or incoherent quotes, we design a data augmentation procedure based on a Beta distribution, the parameters of which are tuned adaptively depending on market data. This procedure can be used even when the considered quote is not corrupted, for the sake of better stability.

\item We perform several tests  on synthetic data, using the SVI model,  and on real data on BTC-USD options; these experiments confirm the gain in accuracy and stability of the procedure. We observe quite systematically an improvement of calibration error by a factor between 25\% and 30\%.
  \end{itemize}

\paragraph{Organization of the paper.}
In Section \ref{section:Market data}, for the sake of completeness and since this is a new emerging market, we provide a short description of the market for options on cryptocurrencies (cryptos for short), along with some comparisons with traditional markets. Notations are also defined in this section. 
In Section \ref{section:Improving calibration} we develop the analysis related to large spreads or missing quotes, and develop the proposed methodology. Numerical tests are presented in Section \ref{section:Numerical tests}. Some technical results are postponed to the Appendix.

\section{Market data}
\label{section:Market data}
We describe the main exchanges for option-type derivatives, as well as some operating characteristics of the contracts. This is a description as of the time of writing, keeping in mind that this is a quickly-evolving market.

\subsection{Market exchanges}
As with decentralized finance, there are numerous option trading platforms, among which we can mention\footnote{\url{https://www.binance.com}, \url{https://www.bybit.com}, \url{https://www.delta.exchange}, \url{https://www.deribit.com},  \url{https://ftx.com}, \url{https://www.huobi.com}, \url{https://www.okx.com}} Binance, Bybit, Delta Exchange, Deribit,  FTX,  Huobi or OKX.
The first platforms appeared in 2016, some of them are more recent.
Most proposed options are European-style options, but Huobi gives access to American-style options.
Almost all  options are vanilla types, i.e. calls and puts; however, Huobi provides  {\it Touch Options Double One-Touch} and {\it Touch Options Double No-Touch}, that are barrier options.

\begin{figure}[htb]
\begin{center}
\includegraphics[scale=0.47]{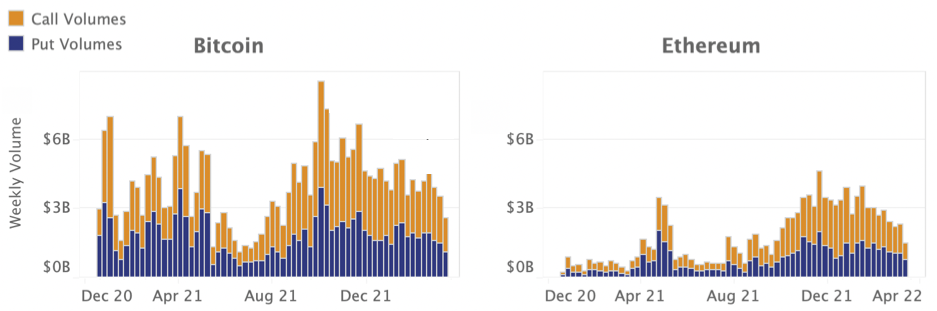}
\caption{Call and put weekly volumes, in billions USD, from December 2020 to April 2022. On the left, the underlying crypto is BTC; on the right, ETH. Source: \url{https://www.kaiko.com}.\label{figure:Bitcoin_and_Ethereum_Option_Volume}}
\end{center}
\end{figure}

The underling asset is usually a currency pair crypto-fiat. The fiat currency (fiat for short) is often USD, although it can be replaced by  USDC\footnote{USDC is a stablecoin pegged to USD. USDC has been developed by two companies, Circle et Coinbase, and it is integrated into Visa payments.} (or any another stablecoin), this is done for instance on Bybit. The usual cryptocurrencies used as  underlyings are BTC or ETH {(or more recently SOL)}. However, we anticipate that other cryptos will also be used in the future, as well as other fiats, such as Japanese Yen (JPY) or South Korean Won (KRW). 
Delta Exchange also offers options on other cryptos, such as BNB, XRP, LINK, AVAX or MATIC.
The payoff of the options can be written in units of crypto or units of fiat, which changes  the pricing rules from a mathematical finance point of view \cite[Chap. 4]{musi:rutk:05}. The settlement usually occurs in fiat.
The global volume of option trading in billions USD is depicted in Figure \ref{figure:Bitcoin_and_Ethereum_Option_Volume}.

\subsection{Some comparative statistics}
Despite the numerous option exchanges, Deribit claims to provide 80\% to 90\% of the global Open Interest. In what follows, we focus the discussion to Deribit contracts, which contain the most common features.
We assume the  reader is  familiar with traditional markets, and we provide a few comparisons to highlight similarities and  main differences with crypto-markets. As a benchmark, we consider options on SP500 on  May 24th 2022, and we compare them with options on BTC-USD the same day.

\paragraph{Maturities/Expiries.} For options on SP500, 43 maturities are listed: 31 in 2022, 9 in 2023, 1 in 2024, 1 in 2025 and 1 in 2026. On BTC-USD, only  8 expiries\footnote{It appears that practitioners dealing with options on cryptos are more likely to refer to expiries rather than maturities.} are available, from  May 27 2022 until  March 31 2023. The scarcity of expiries for cryptocurrency options is certainly a factor in increasing the liquidity of available contracts by concentrating trading on these contracts. On the other hand, this scarcity makes it difficult to construct a full implied volatility surface, because of the lack of available expiries in the term-structure of \IV s.

\paragraph{Log-forward-moneyness.}
In Tables \ref{table:SP500} and \ref{table:BTCUSD}, we list the strikes and the related log-forward-money\-nes\-ses for the available quotes for options. Observe that the width of available log-forward-moneynesses for both underlyings is quite similar, regardless of the expiry. The main difference comes from the center of the log-forward-moneyness range: it is biased to the left for SP500, showing the interest for options with small strikes (to help asset manager hedge themselves against market crashes), and it is biased to the right for BTC-USD, probably because the market has  often been bullish.
\begin{table}[htb]
  \centering
    \begin{tabular}{cccc}
    \toprule
    Maturity  & Strike (quotes)  & Log-forward-moneyness  & Length range \\
       & (min, max) &  range (quotes) & (quotes) \\
    \otoprule
    1 D (2022-05-25)  & [ 1000, 5200 ] & [ -0.5991, 0.1168 ] & 0.7159 \\
    \midrule
    1 W (2022-06-01) & [ 1000, 5000 ] & [ -0.5991, 0.0998 ] & 0.6989 \\
    \midrule
    1 M (2022-06-24) & [ 1000, 5600 ] & [ -0.5991, 0.1490 ] & 0.7481 \\
    \midrule
    3.5 M (2022-09-16) & [ \phantom{1}200, 7400 ] & [ -1.2980, 0.2701 ] & 1.5681 \\
    \midrule
    7 M  (2022-12-16)   & [ \phantom{1}100, 7300 ] & [ -1.5991, 0.2642 ] & 1.8633 \\
    \midrule
    1 Y (2023-06-16) & [ \phantom{1}200, 7400 ] & [ -1.2980, 0.2701 ] & 1.5681 \\
    \midrule
    2 Y (2024-12-20) & [ \phantom{1}200, 9200 ] & [ -1.2980, 0.3646 ] & 1.6626 \\
    \bottomrule
    \end{tabular}
  \caption{SP500: available log-moneyness on quotes, at 2022 May 24th, for a few maturities}
  \label{table:SP500}
\end{table}

\begin{table}[htb]
  \centering
    \begin{tabular}{cccc}
    \toprule
    Expiry   & Strike (quotes)  & Log-moneyness  & Length range \\
       & (min, max) &  range (quotes) & (quotes) \\
        \midrule
    1 M  (2022-06-24) & [ 10000, \phantom{1}75000 ] & [ -0.4655, 0.4095 ] & 0.875 \\
            \midrule
    3.5 M (2022-09-30) & [ 10000, 200000 ] & [ -0.4655, 0.8355 ] & 1.301 \\
                \midrule
    7 M  (2022-12-30) & [ 10000, 300000 ] & [ -0.4655, 1.0116 ] & 1.4771 \\
                    \midrule
     11M (2023-03-31) & [ 10000, 300000 ]  & [ -0.4655, 1.0116 ] & 1.4771 \\
                        \bottomrule
    \end{tabular}
  \caption{BTC-USD: available log-moneyness on quotes, at 2022 May 24th, for a few expiries.}
  \label{table:BTCUSD}
\end{table}

\paragraph{Tick size.}  As a final comparison, 
we consider the tick size on option contracts, and compare it with different underlying assets, ranging from SP500 to EUR-USD, see Table \ref{table:compare:ticksize}.
\begin{table}[htb]
    \begin{center}
    \begin{tabular}{cccccc}
        \toprule Option $\backslash$ Exchange & CBOE & EUREX & EURONEXT  & CME & Deribit   \\
        \otoprule
        Underlying Name & SP500 & Eurostoxx & AEX & EUR-USD & BTC-USD \\
        Option tick size & 0.05 & 0.10 & 0.01  & 0.00005 & 0.0005\\
        Underlying Value (close) & \multirow{2}{*}{3 941.48}  & \multirow{2}{*}{3 647.56	} & \multirow{2}{*}{679.88} & \multirow{2}{*}{1.0737} & \multirow{2}{*}{1}\\
        on 2022, May 24th & & & & & \\
        \midrule
        Ratio in basis points  & \multirow{2}{*}{0.127} & \multirow{2}{*}{0.274} & \multirow{2}{*}{0.147}   & \multirow{2}{*}{0.466} &  \multirow{2}{*}{5} \\
        (bps, i.e. 1E-4) &  &  &    &  &   \\
        \bottomrule
    \end{tabular}
    \caption{Comparison of tick sizes for option prices written on various underlyings on different exchanges. On Deribit, the tick size is expressed in BTC. 
    \label{table:compare:ticksize}}
    \end{center}
\end{table}
Compared to traditional markets, the relative tick size (defined as the ratio between the option tick size and the underlying asset value) is much larger for options on cryptos, by a factor ranging between 10 and 50. As a result, the relative  spread must be significantly higher and hence, inferring the efficient price as the mid-price entails larger errors,  that undeniably impact the accuracy and stability of the calibration.

\begin{figure}[H]
    \begin{center}
        \includegraphics[scale=0.38]{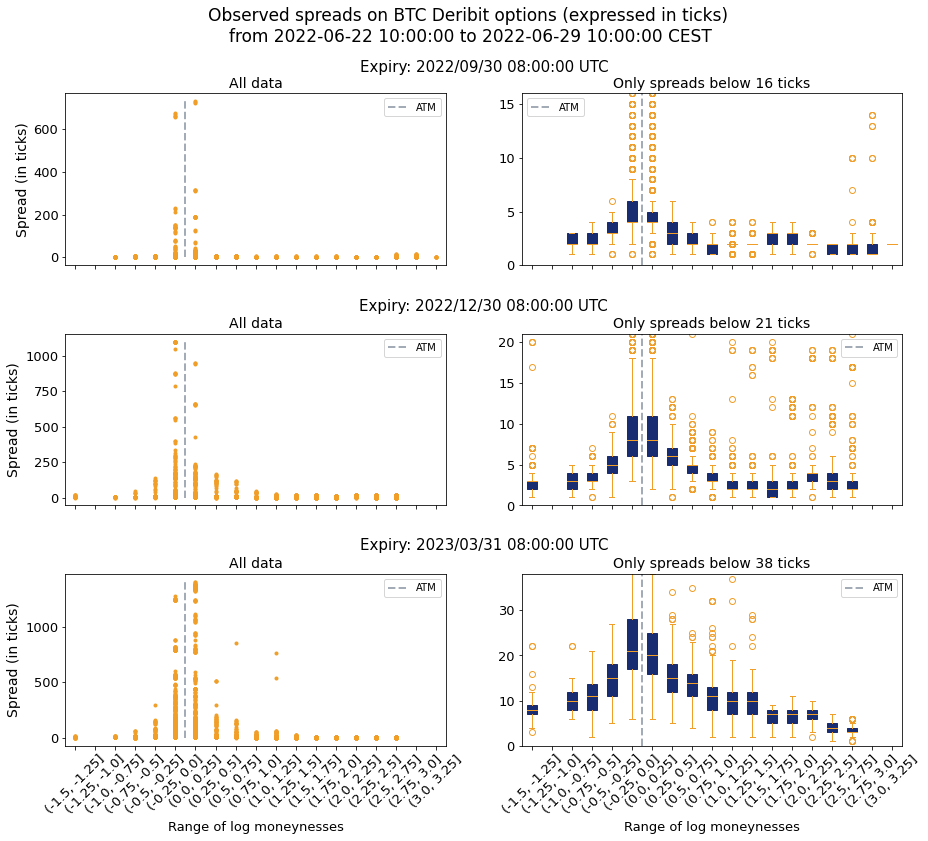}
        \caption{Boxplot of spreads (expressed in ticks) for data collected over a period of a week (2022-06-22 10:00 to 2022-06-28, 10:00 CEST). Option expiries are: 2022 Sept 30th (top), 2022 Dec 30th (middle), 2023 March 31th (bottom). The $x$-axis corresponds to observed log-moneynesses, the data points are those from BTC-USD options quoted on Deribit.
        On the left: all data; on the right: data with spreads below a threshold (respectively 16 ticks, 21 ticks, 39 ticks). }
        \label{figure:spead:ticksize}
    \end{center}
\end{figure}

For the sake of completeness, we show in Figure \ref{figure:spead:ticksize} the boxplots of spreads over a period of 1 week, for different expiries. Observe that extreme spreads can occur (over 100 ticks). For extreme log-moneynesses, the spread is usually 1 or 2 ticks for the shortest expiry, and up to 10 ticks for longer expiries; at the money, the spreads are often of order 5 to 15 ticks, depending on the expiries. This highlights the needs for calibration methodologies that are accurate and robust even when the spreads are (extremely) large. The calibration process we describe in the upcoming sections improves the one that simply takes mid-prices as inputs.

In addition, we note that on average, 10\% of the quotes have no bid price and 5\% of the bid prices are at 1 tick. This type of missing or out-of-arbitrage-bounds quotes prevent us from obtaining accurate and stable results using the naive mid-price in the calibration process. This problem is addressed with a data augmentation procedure in Section \ref{subsection:Data augmentation when missing data}.
\label{section:missing_bid}

\subsection{Quantitative modeling}
We fix a few notations, used in all the sequel for the quantitative analysis.

\begin{itemize} 
\item We consider two currencies: one for the "Fiat economy" with prices in units of \F~(typically  \USD), and one for the "Crypto economy" with prices in units of \C~(typically \BTC,  \ETH).
\item $X$ stands for the crypto-fiat rate: 1 unit of \C~is worth $X$ units of \F. 
\item We denote the prices of zero-coupon bonds by
$\BF_{t,T}$ and $\BC_{t,T}$, respectively in  the  fiat and crypto currencies.
\item We denote by $\CF_t(\Phi^\F_T,T)$ and $\CC_t(\Phi^\C_T,T)$ the cash-prices in \F~and \C~currencies at time $t$, for some  cashflows $\Phi^\F_T$ or $\Phi^\C_T$ delivered at time $T$.
\end{itemize}

The no-arbitrage condition between currencies \C~and \F~yields the relation
  \begin{align}
\label{eq:NA:relation:2}
\frac1{X_{t}}\CF_t(\Phi^\F_T,T)=\CC_t\left(\frac{\Phi^\F_T}{X_{T}},T\right),  \end{align}
meaning that it must be equivalent to compute the \F-price at time $t$ of payoff $\Phi^\F_T$ at expiry $T$, then convert it in \C; and to  compute directly the \C-price at time $t$ of payoff $\Phi^\F_T/X_T$ at expiry $T$.

\subsubsection{Options}\label{subsubsection:deribit:options}
According to \textit{Deribit - Options Documentation}\footnote{\url{https://legacy.deribit.com/pages/docs/options}}, call and put options are European-style options, written on the crypto-fiat rate $X$. The options are priced in \C.

\begin{example}[from Deribit] A trader buys a call option with a strike price of 10,000  \USD~for 0.05 BTC. Now they have the right to buy 1 \BTC~for 10,000  \USD.
\begin{itemize}
\item At expiry, the \BTC~Index is at 12,500  \USD, and the delivery price is 12,500  \USD.

In this case, the option is settled for 2,500  \USD~per 1 \BTC. At expiry, the trader's account is credited with 0.2 \BTC~(2,500/12,500), and the seller's account is debited with 0.2 \BTC. The initial purchase price was 0.05 \BTC; therefore, the trader's profit is 0.15 \BTC.

\item Any call option with an exercise price (strike price) above 12,500  \USD~will expire worthless.
\end{itemize}
This example shows that the payoff (in \C) of a call option on the exchange rate \C-\F~with strike $K$ is therefore
\begin{equation}
    \label{eq:payoff:call}
    \frac{(X_T-K)_+}{X_T}.
\end{equation}
\end{example}
In view of \eqref{eq:NA:relation:2}, the Deribit call price is 
\begin{align}\label{eq:prix:call:deribit}
\CallDer_{t}(T,K)= \CC_t\left(\frac{(X_T-K)_+}{X_T},T\right)=
\frac1{X_{t}}\CF_t((X_T-K)_+,T).
 \end{align}

\subsubsection{Futures}
We refer to the \textit{Deribit - Futures Documentation}\footnote{\url{https://legacy.deribit.com/pages/docs/futures}} to describe its so-called Future contract. Actually, Deribit is not a usual central clearing counterparty (CCP) because there are no daily margin calls. Hence, a Future in the Deribit terminology is not a Future in the traditional sense and the pricing rule must be clarified. 

Expirations take place on Fridays at 08:00 UTC for weekly Futures, and on the last Friday of the month for monthly and quarterly Futures. In all cases there is a Cash settlement in \C. The value of the Deribit Future $\FutDerT_{t}$ at time $t$ for expiry $T$ is 	1  \USD~per Index Point, with contract size  \USD~10 (on \BTC) and  \USD~1 (on  \ETH); let us denote generically this size by $\Size$.

\begin{example}[adapted from Deribit] A trader buys 100 futures contracts on the \BTC~index, at $\FutDerT_{t}=10,000$  \USD~per \BTC. The trader is now long   1,000  \USD~worth of \BTC~with a price of 10,000  \USD~(100 contracts $\times$ 10  \USD~= 1,000  \USD).
\begin{itemize}
\item At inception, the trader does not pay anything, but is committed to paying 1,000  \USD~at expiry; this amount is worth $1,000\ \BF_{t,T}$ \USD ~at inception.
\item At expiry, the trader receives $1,000/10,000$ \BTC, because of the futures contract.
\item They can sell this amount in \BTC~at the expiry rate $X_{T}$; for instance if $X_{T}=12,000$  \USD, then this amount is worth $\frac{1,000}{10,000} \times 12000= 1200$ \USD.
 \end{itemize}
 \end{example}
This example (Reverse Cash-and-Carry Arbitrage) shows that the payoff of this strategy at expiry is 
$$\frac{100\ \Size}{\FutDer_{t}}X_{T}\ \USD.$$
Equating the \F-values, we get
\begin{align} \label{eq:future:X}
100\ \Size\ \BF_{t,T}=\CF_{t}\left(\frac{100\ \Size}{\FutDerT_{t}}X_{T},T\right)\Longleftrightarrow \FutDerT_{t}=\frac{\CF_{t}(X_{T},T)}{\BF_{t,T}}.
 \end{align}
This shows that the Deribit  Future contract is like an {\it OTC-forward contract}. Invoking again the no-arbitrage relation  \eqref{eq:NA:relation:2}, we get
\begin{align}
\label{eq:NA:relation:1:ctd}
\FutDerT_{t}=\frac{X_{t}\CC_t(1,T)}{\BF_{t,T}}=X_{t}\frac{\BC_{t,T}}{\BF_{t,T}}.
 \end{align}
This relationship between forward  and spot exchange rates on the one hand, and interest rate products in \C~and \F~economies on the other hand, is known as Interest Rate Parity (IRP) relationship \cite[p.153]{musi:rutk:05}.

\subsubsection{Implied volatility}
We are now ready to define the implied volatility of the Deribit call price with expiry $T$ and strike $K$. We first recall the application  of the Black formula \cite{gema:elka:roch:95}, under the assumption that the \textit{forward} $\CF_{t}(X_{T},T)/\BF_{t,T}=\FutDer_{t}$ admits a deterministic volatility $\sigma^\Fut_{.}$: the call price in \F\ is then
 \begin{align}
    \Call^\F_t(T,K)
    =\ & \BF_{t,T}\left(\FutDer_{t} \,\cN\left(d_+\right)-K\,\cN\left(d_-\right)\right),    \label{eq:prix:call:on:X:future:trading}\\
\text{with}\qquad    d_\pm&=\ \frac{1}{\sqrt{\int_t^T|\sigma^\Fut_u|^2 \du} }\log\left(\frac{\FutDer_{t}}{K }\right)\pm\frac{1}{ 2}\sqrt{\int_t^T|\sigma^\Fut_u|^2 \du}.\nonumber
  \end{align}
  This is simply an extension of the Garman-Kohlhagen formula \cite{garm:kohl:83} with possibly stochastic interest rates. Combining the above and  \eqref{eq:prix:call:deribit}, we define the implied volatility \IV~of the Deribit call price as the constant parameter in \eqref{eq:prix:call:on:X:future:trading} that matches market prices.
  
  \begin{definition} The implied volatility \IV~at time $t$ for expiry $T$ and strike $K$ is given implicitly by
\begin{align}\label{eq:def:IV} 
\frac{X_{t}\CallDer_{t}(T,K)}{\BF_{t,T}}&\ \isdef\ \FutDerT_{t} \,\cN\left(d_+\right)-K\,\cN\left(d_-\right)\\
\text{with}\qquad    d_\pm&\ =\ \frac{1}{\IV\sqrt{\tau} }\log\left(\frac{\FutDerT_{t}}{K }\right)\pm\frac{1}{ 2}\IV\sqrt{\tau}, \qquad \tau=T-t.\nonumber
 \end{align}
  \end{definition}  
The implied volatility can be retrieved similarly from put prices, with obvious adjustments to the formulas.

\section{Improving calibration}
\label{section:Improving calibration}
 {In the following, we assume that calibration is performed at the current time, i.e.,  the time $t$ is fixed at 0, and we omit to indicate this time dependency in the upcoming notations.
To emphasize the dependency of the implied volatility \eqref{eq:def:IV} w.r.t. expiry and strike, we specialize notations:
\begin{itemize}
\item $\wz\isdef\tau\cdot\IV^2(\tau,k)$ for the implied total variance at time $0$;
\item $k\isdef\log(\frac{K}{\FutDerT_{0}})$ for the log-forward-moneyness;
\item $\tau\isdef T-t=T$  for the time-to-expiry;
\item whenever the dependency w.r.t. the price $P$ is important, we will write $\IV(P,\tau,k)$ instead of $\IV(\tau,k)$.
 \end{itemize}

 The bid and ask prices for log-forward-moneyness $k$ and time-to-expiry $\tau$ are respectively denoted by $\Bid(\tau,k)$ and $\Ask(\tau,k)$. We write 
 \begin{align}
 \md(\tau,k)\isdef \frac{1}{2 }\Big(\Bid(\tau,k)+ \Ask(\tau,k)\Big)\quad \text{ and }\quad \spr(\tau,k)\isdef\Ask(\tau,k)-\Bid(\tau,k),
 \end{align}
 for the mid-price and the spread respectively. The notation $\ts$ stands for the tick size value on options.}

\subsection{Filtering OTM options in the data set}
Thanks to the call-put parity relationship, if we assume a zero spread on the call and put prices, then the implied volatility is the same for both products, provided  they have the same expiry and strike. However, when there is a non-zero  spread, the implied volatilities will differ  and we investigate below the way the choice of the option type (put or call) impacts the computation of implied volatilities. 
The larger the tick size, the larger the spread, and the larger the impact on the computed  implied volatility.

The impact analysis can be visualized using a sensitivity indicator, when the price is bumped by the relative error $\varepsilon$, i.e.
\begin{align*}
\text {Initial price}\cdot (1+\varepsilon)&= \text {Black-Scholes price}(\IV^\varepsilon(\tau,k)),
\end{align*}
for some bumped  implied volatility $\IV^\varepsilon(\tau,k)$. The  parameter $\varepsilon$ encodes the fact that the efficient price is not observed exactly but with an error. Alternatively to a relative sensitivity, one could have considered an absolute sensitivity but we believe that reasoning with relative changes makes the argumentation more intuitive. 

By differentiating w.r.t. $\varepsilon$, we get
\begin{align*}
\text{relative \IV\ sensitivity}& \isdef \partial_\varepsilon \IV^\varepsilon\Big|_{\varepsilon=0}= \ \frac{\text {Initial price}}{\vega^\BS(\IV^\varepsilon)}\Big|_{\varepsilon=0}.
\end{align*}

\begin{figure}[H]
\centering
\includegraphics[scale=0.43]{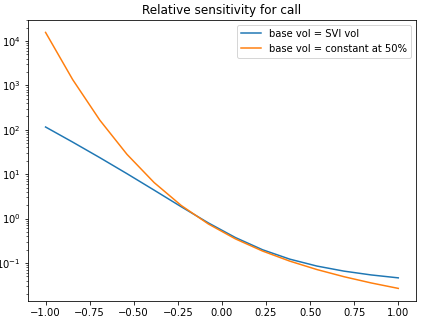}\qquad\qquad
\includegraphics[scale=0.43]{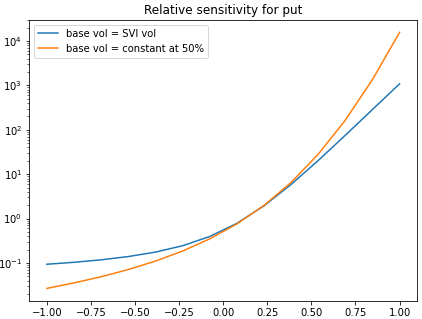}

\caption{Sensitivity $\partial_\varepsilon \IV^\varepsilon\big|_{\varepsilon=0}$ as a function of the log-forward-moneyness $k$, under different \IV\ models:  SVI and Constant volatility. 
Relative sensitivities for calls are depicted on the left, those for puts on the right. \label{fig:sensitivity:IV}}
\end{figure}

The above sensitivity depends in particular on the log-forward-moneyness. We depict it on Figure \ref{fig:sensitivity:IV}  under two working assumptions on volatilities: one is computed with an SVI model with parameters $(a,b,\rho, m,\sigma)=(0.02, 0.2, -0.3, 0.05,0.6)$ (see Equation \eqref{eq:raw:par} below), and the other with a constant volatility equal to 50\%. The other parameters are $\tau=0.25$ and $\Fut=1$. The values in the plots depend on the considered \IV~but the shape globally remains the same.

Figure \ref{fig:sensitivity:IV} reads as follows: a relative error on market call price of 10\% ($\varepsilon=0.1$) entails an absolute 
\IV\ error that ranges from  $0.1\times 10^2\times 10^4=100\, 000$ {\bps} to $0.1\times 10^4\times 10^4=10 \,000\, 000$ \bps~for ITM options ($k\ll0$), and that is approximately $0.1\times10^{-1}\times 10^4=100$ {\bps} for OTM options ($k\gg0$). The situation is symmetric for puts. All in all, the relative sensitivity is much smaller for OTM options than for ITM options. 

This is the reason why in our calibration procedures, we filter bid-ask prices by keeping only OTM options, i.e. we only consider call prices (resp. put prices) for $k\geq 0$ (resp. $k\leq 0$).

\subsection{Bias on the calibration}
{In this section we evidence that the calibration of the implied total  variance from noisy observed prices systematically induces an underestimation of the efficient total variance in the left and right wings. The discussion is based on a statistical learning point-of-view. The removal of the bias is discussed in the next section.  

To formalize the bias result, we assume that the mid-price is a centered-noisy estimate of the efficient price, which is a standard and reasonable assumption from a microstructural point of view:

\begin{enumerate} 
\item [{\bf (H-1)}] \label{ass:1} For any $(\tau,k)$, the mid-price is an unbiased estimate of the efficient price $\Eff$:
\begin{align}
\md(\tau,k)=\Eff(\tau,k)+\varepsilon(\tau,k),
\label{eq:midprice}
\end{align}
where $\varepsilon(\tau,k)$ is not identically zero and is centered.
\end{enumerate}

This  assumption is fulfilled for instance by the  model in \cite{rose:09}, where the author assumes  the efficient price data is observed with a
round-off error, which leads to an observation noise $\varepsilon(\tau,k)$ satisfying \ref{ass:1}. Further details on this modeling point-of-view are provided in Section \ref{subsection:Augmenting data in the bid-ask interval}. The fact that $\varepsilon(\tau,k)$ is not identically zero is a technical condition for Theorem \ref{th:biais}, it is obviously satisfied in practice.
Here the underlying probability space $(\Omega, \P, {\cal F})$  accounts for the randomness of the noise $\varepsilon(\tau,k)$ for all $\tau, k$, and the efficient prices $\setof{\Eff(\tau,k)}{\tau,k}$ are considered as fixed (calibration is performed at time $t=0$).

Under the sole Assumption \ref{ass:1}, we now justify that the calibration is strongly biased.
 As a calibration problem, we aim at fitting a parametric model on the efficient $\IV$ (i.e., the one related to efficient prices), by solving a calibration problem on $n$ mid-price data points:
\begin{align}\label{eq:calibration:pb}
\inf_\chi \frac{1}{ n} \sum_{i=1}^n\Big(\tau \cdot \IV^2(\md(\tau,k_i),\tau,k_i)-\omega(\tau,k_i;\chi)\Big)^2, 
\end{align}
where $\omega(\tau,k_i;\chi)$ is a given total variance parametrization (for some parameters $\chi$) at the log-forward-moneyness $k_i$ for a given observed data point. The goal is to obtain the best parameters $\chi^\star$ for which 
\begin{align}
\label{eq:objective}
\omega(\tau,k; \chi^\star)\approx \tau \cdot \IV^2(\Eff(\tau,k),\tau,k),{\quad\text{for all $k$}}.
\end{align} In the subsequent experiments of Section \ref{section:Numerical tests}, we consider  the SVI parametrization \eqref{eq:raw:par} for the total variance parametrization  $\omega$.

 The above problem is similar to a regression problem \cite{gyor:kohl:krzy:walk:02} that consists in  finding a function $\mathcal M$ such that $Y={\cal M}(X)+\textrm{noise}$, given the observation of an $n$-sample of $(Y,X)$ called "Response, Design". 
 The main difference is that in \eqref{eq:calibration:pb}, we do not minimize the squared difference between $Y$ and functions of $X$, but between a nonlinear function of $Y$ (here $\tau \cdot\IV^2(\cdot)$) and functions of $X$. This composition with a non-linear function changes the theoretical analysis.

To summarize, problem \eqref{eq:calibration:pb} is an optimization problem over a finite-sample data, which can be compared to the ideal optimization problem
\begin{align}\label{eq:calibration:pb:esp}
&\inf_\chi\int \Esp{\Big(\tau \cdot \IV^2(\md(\tau,k),\tau,k)-\omega(\tau,k;\chi)\Big)^2}\pi(\dd k),
\end{align}
where $\pi$ is the distribution of the observed log-forward-moneyness. Denote by $\varphi$ the squared function within the above expectation. Then the difference between  \eqref{eq:calibration:pb} and \eqref{eq:calibration:pb:esp} is bounded by the so-called supremum of empirical process 
\begin{align}\label{eq:calibration:pb:esp:sup}
\sup_\chi\left(\frac{1}{n }\sum_{i=1}^n \varphi(\md(\tau,k_i),\tau,k_i;\chi) -\int \Esp{\varphi(\md(\tau,k),\tau,k;\chi)}\pi(\dd k)\right).
\end{align}
The above statistical error can be quantified depending on the assumptions on the tails of the quantity $\varphi(\md(\tau,k),\tau,k;\chi)$: under boundedness conditions, see \cite{klei:rio:05}; or under light and heavy tail conditions, see \cite{cham:gobe:szab:20} and \cite{cham:gobe:liu:21}. See also \cite{kolt:11} for a broad overview of error bounds in statistical learning. Essentially, these results state that the probability of 
\eqref{eq:calibration:pb:esp:sup} being greater than $C \delta/\sqrt n$  is exponentially small in $\delta >0$: in other words, the  minimum in \eqref{eq:calibration:pb} is close to \eqref{eq:calibration:pb:esp} with high probability.

Furthermore, a bias-variance decomposition of \eqref{eq:calibration:pb:esp} allows to write
\begin{align}\nonumber
\inf_\chi\int \left[\Var{\tau \cdot \IV^2(\md(\tau,k),\tau,k)} +
\Big(\Esp{ \tau \cdot \IV^2(\md(\tau,k),\tau,k)-\omega(\tau, k;\chi)}\Big)^2\right] \pi(\dd k).
\end{align}
Since the first term does not depend on the parameter $\chi$, the calibration algorithm will take  mid-prices as inputs and attempt to minimize the integral of the squared bias error, i.e., to compute 
\begin{align}\label{eq:calibration:pb:bias}
\inf_\chi\int\Big(\Esp{ \tau \cdot \IV^2(\md(\tau,k),\tau,k)-\omega(\tau,k;\chi)}\Big)^2\pi(\dd k).
\end{align}
\begin{theorem}\label{th:biais}
Assume \ref{ass:1} holds.  In the wings (i.e., when $|k|\gg0$), the calibration algorithm with mid-prices as inputs calibrates a negatively-biased  total implied variance, while at the money ($k\approx 0$), the  calibration algorithm   calibrates a positively-biased  total implied variance.

The above under-estimation in the wings is similar if the mid-price error
$$\left(\tau \cdot \IV^2(\md(\tau,k_i),\tau,k_i)-\omega(\tau,k_i;\chi)\right)^2$$
in \eqref{eq:calibration:pb} is replaced by the average over bid and ask prices
$$\frac{1}{ 2}\left(\tau \cdot \IV^2(\Bid(\tau,k_i),\tau,k_i)-\omega(\tau,k_i;\chi)\right)^2
+\frac{1}{ 2}\left(\tau \cdot \IV^2(\Ask(\tau,k_i),\tau,k_i)-\omega(\tau,k_i;\chi)\right)^2
.$$

\end{theorem}
\begin{proof} From Lemma \ref{lemma:IV:derivative} in Appendix \ref{section:appendice},  the function $P\mapsto \tau\cdot \IV^2(P)$ is strictly concave when $|k|$ is large enough, and strictly convex for small values of $|k|$, see Figure \ref{fig:convexity:IVsquare}. 
\begin{figure}[t]
\begin{center}
\includegraphics[width=0.24\textwidth]{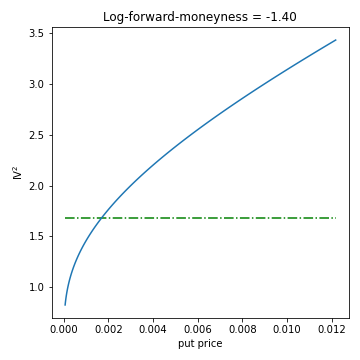}
\includegraphics[width=0.24\textwidth]{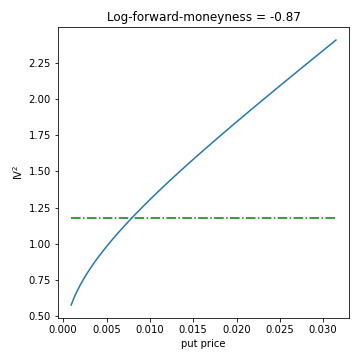}
\includegraphics[width=0.24\textwidth]{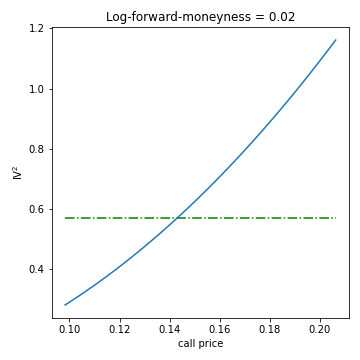}
\includegraphics[width=0.24\textwidth]{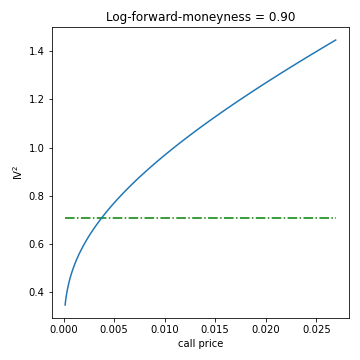}
\caption{Visualization of the concavity (in the wings) or convexity (ATM) of the function $P\mapsto  \IV^2(P)$. The underlying contract is a call (resp. put) for positive (resp. negative) log-forward-moneyness. From the left to the right: $k=-1.40, k=-0.87, k=0.02, k=0.90$. The dashed line corresponds to the volatility in an SVI model as in Figure \ref{fig:raw:svi}, and with $\tau=0.25$.
  \label{fig:convexity:IVsquare}}
\end{center}
\end{figure}

Because $\varepsilon(\tau,k)$ is not identically zero, the Jensen inequality becomes a strict inequality and writes for log-forward-moneyness in the wings ($|k|\gg0$):

\begin{align}
\tau \cdot \IV^2(\Eff(\tau,k),\tau,k)
\underset{\ref{ass:1}}{=} \tau \cdot \IV^2(\Esp{\md(\tau,k)},\tau,k)
 >  \Esp{ \tau \cdot \IV^2(\md(\tau,k),\tau,k)}.
\label{eq:jensen:1}
 \end{align}
 The reverse inequality holds when $k\approx 0$. This shows that in the minimization problem \eqref{eq:calibration:pb:bias}, the parametrization $\omega(\tau,k;\chi)$ attempts to fit a value below (resp. above) the expected one (which is $\tau \cdot \IV^2(\Eff(\tau,k),\tau,k)$) in the wings (resp. at the money). This proves the first theorem statement.
 
When the mid-price is replaced by the average of bid and ask prices, the analysis is similar and shows that the bias is even higher in the wings (when $|k|\gg0$). Indeed,
\begin{align}
\tau \cdot \IV^2(\md(\tau,k),\tau,k)
&=\tau \cdot \IV^2(\frac{1}{2}\Bid(\tau,k)+\frac{1}{2}\Ask(\tau,k),\tau,k)\\
&>\frac{1}{2}\tau \cdot \IV^2(\Bid(\tau,k),\tau,k)
+\frac 12 \tau \cdot \IV^2(\Ask(\tau,k),\tau,k),
 \end{align}
{which is a larger inequality than that of \eqref{eq:jensen:1}}.
\end{proof}
Theorem \ref{th:biais} shows that any parametric form $\omega(\tau,k;\chi)$ within the calibration procedure will lead to an underestimation of \IV~in the wings and an overestimation at the money. Presumably, this phenomenon might be mitigated by the additional approximation error due to the parametric form, but nonetheless,  the overall behavior should remain the same. 
\begin{figure}[t]
\begin{center}
\includegraphics[width=0.19\textwidth]{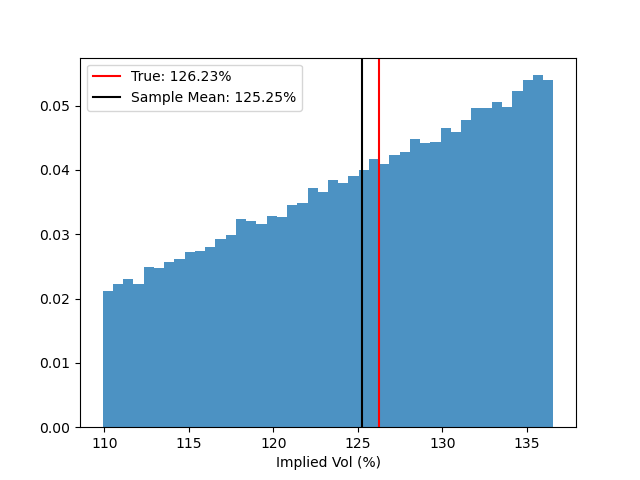}
\includegraphics[width=0.19\textwidth]
{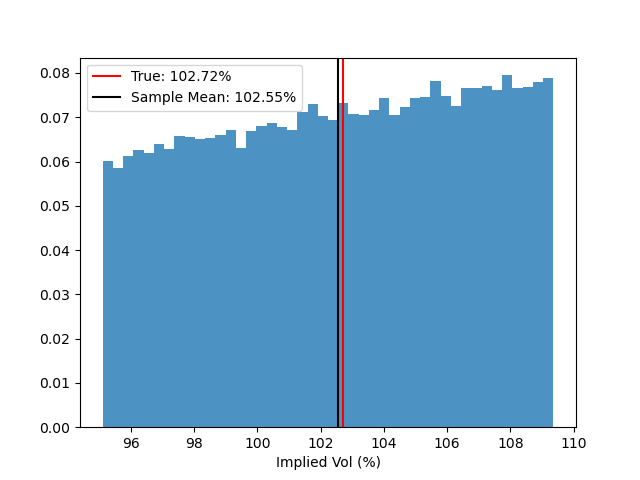}
\includegraphics[width=0.19\textwidth]{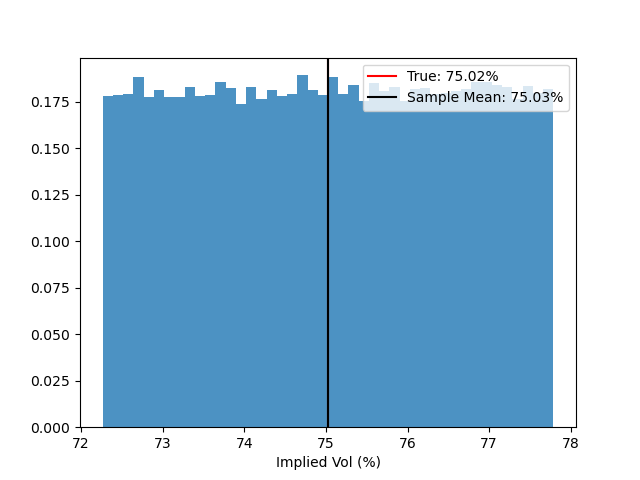}
\includegraphics[width=0.19\textwidth]{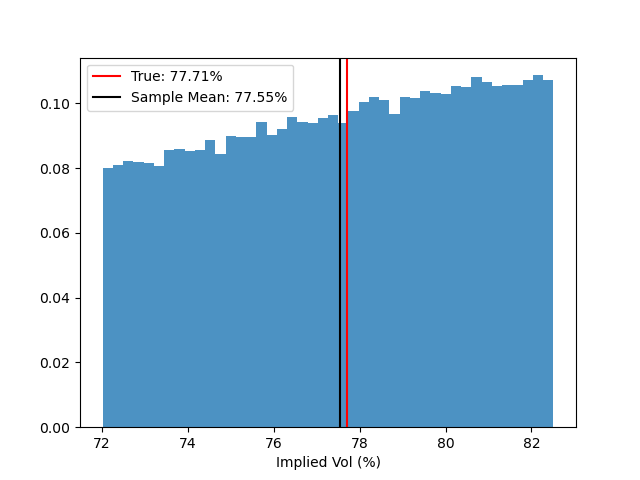}
\includegraphics[width=0.19\textwidth]{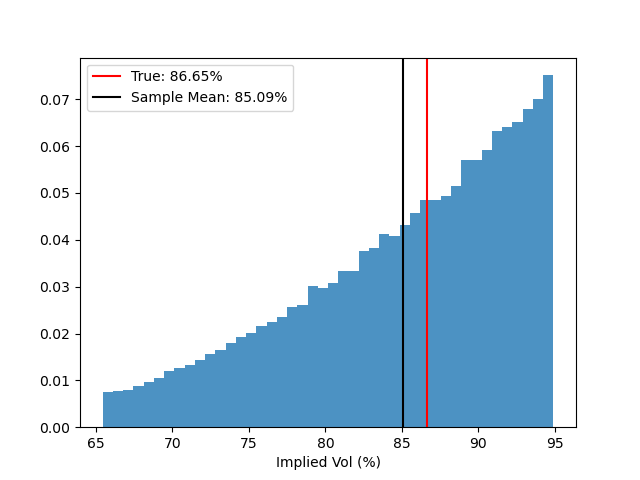}

\caption{Distribution over 10 000 samples of the mid-price IV in the model \ref{ass:2}, when the effective price is given by  an SVI smile, with the same parameters as in Figure  \ref{fig:raw:svi}. The spread is set to 4 ticks in the wings, and 20 ticks at the money. Colors: the exact IV is in red; the IV mean over the samples is in black. Values of the log-forward-moneyness $k$ from the left to the right: $k=-1.31, k=-0.73, k=0.04, k=0.62, k=1$. \label{fig:bias:viz}
}
\end{center}
\end{figure}

On traditional markets where the spread is small, the above over/under-estimation is definitely negligible but for options on cryptos for which the relative tick size is larger (see Table \ref{table:compare:ticksize} and Figure \ref{figure:spead:ticksize}), the bias  becomes a more significant concern. A visualization of the calibration bias from Theorem \ref{th:biais} is depicted in Figures \ref{fig:bias:viz} and \ref{fig:recalibration}, where we use synthetic data generated by an SVI model that gives the so-called efficient price, to which we add a noise according to model \ref{ass:2} to get the mid-price on which we calibrate. Consistently with Lemma \ref{lemma:IV:derivative} which shows that $\partial_P (\IV^2)$ goes to $+\infty$ as $|k|$ gets larger all other things being equal, the bid and ask prices result in an increasing spread on {\IV}s as $|k|$ increases (see the $x$-ranges in Figure \ref{fig:bias:viz}). Figure \ref{fig:bias:viz} evidences that as $|k|$ gets larger, the mid IV (in black) increasingly underestimates the true IV (in red) -- the IV bias can be larger than 100 bps -- while at the money, the bias remains small, and even negligible. In Figure \ref{fig:recalibration}, an SVI model is calibrated on the mean of the mid total variance (according to model \ref{ass:2} below): in other words, this implements criterion \eqref{eq:calibration:pb:bias} which is the so-called square bias term in the calibration. Since the mid IV underestimates in mean the true IV, the calibrated IV curve underestimates the true one in the wings, as predicted by Theorem \ref{th:biais}.

\begin{figure}[t]
\begin{center}
\includegraphics[scale=0.25]{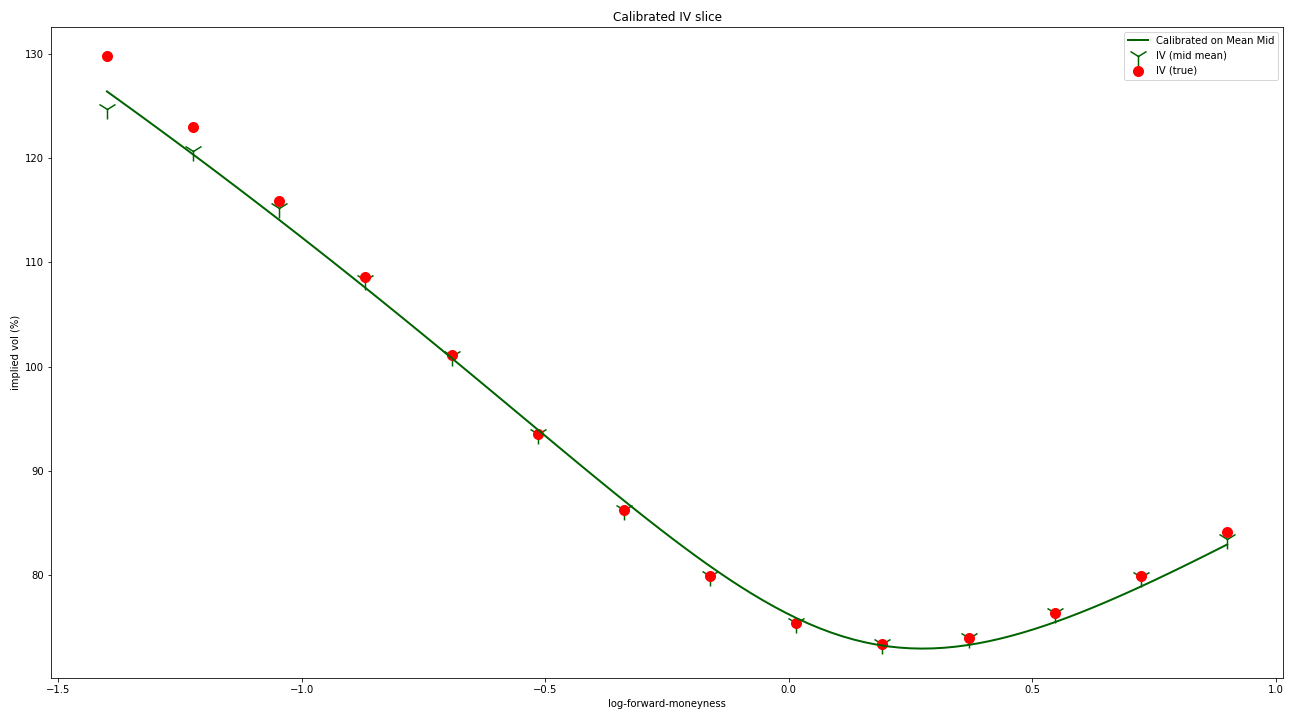}
\caption{Recalibration of a  SVI smile with the same parameters as in Figure \ref{fig:raw:svi}. 
The spread is set to 10 ticks for any log-forward-moneyness. The calibration data comes from the mean of mid total variance (the related mean mid IV is depicted) where the mid-price is generated according to \ref{ass:2}.
\label{fig:recalibration}}
\end{center}
\end{figure}
}

\subsection{Getting rid of  the mid-price bias}
\label{subsection:Augmenting data in the bid-ask interval}
The previous analysis shows that taking  mid-prices as inputs of the calibration procedure induces a systematic bias, which is not corrected by taking the bid and the ask prices with equal probabilities. To compensate for this, we develop a novel approach that uses all the information from the bid and ask interval. This approach is based on determining an anchor price that depends on the bid and ask prices, and ideally leads to an unbiased estimator of the IV corresponding to efficient prices (called the \textit{efficient IV} in what follows).

\paragraph{Anchor price.}
Our analysis is based on a probabilistic model of the efficient price, the mid-price and the spread.
\begin{enumerate} 
\item [{\bf (H-2)}] \label{ass:2} The noise $\varepsilon(\tau,k)$ in \ref{ass:1} has a symmetric distribution, uniformly distributed on an interval with a size equal to the spread; more formally,
\begin{align}
\md(\tau,k)=\Eff(\tau,k)+\spr(\tau,k)\cdot  U, 
\label{eq:midprice=model}
\end{align}
where $U$ is uniformly distributed on $[-1/2,1/2]$, and independent from  $\spr(\tau,k)$.
\end{enumerate}
Here again, we assume that $\Eff(\tau,k)$ is fixed; the spread $\spr(\tau,k)$ on the other hand can be random.

\begin{figure}[H]
\begin{center}
\includegraphics[scale=0.45]{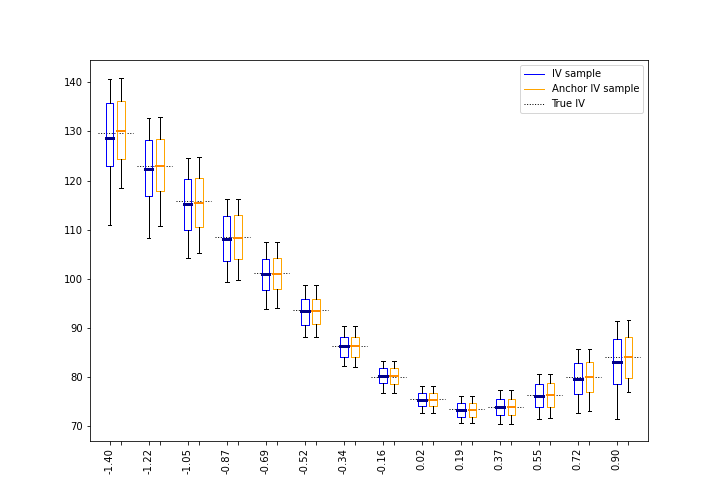}
\caption{Boxplots, for different log-forward-moneynesses on $x$-axis, of implied volatilities based on mid-price or anchor price. Size of the sample: 100 000. The solid lines in the boxplots correspond to the empirical mean over the sample. The parameters for the efficient price and the spread are the same as in Figure \ref{fig:raw:svi}.\label{fig:boxplot:anchor}}
\end{center}
\end{figure}

This model is inspired by \cite{rose:09}, which assumes the efficient price data is observed with a round-off error. 
The model derivation is as follows, where we may omit the index $\tau,k$ for the sake of clarity.  From Lemma \ref{lemma:round:off:error}, and assuming for the time being that the spread is one tick (in other words, that $\delta = 1$), we can assert that the efficient price is related to the bid by the relation 
\begin{align}
\Eff-\Bid\isdef V \cdot  \delta
\end{align}
where $V$ is asymptotically (as $\delta\to 0$) uniformly distributed on $[0,1]$. Using again that the spread is taken to be one  tick, we easily get
\begin{align}
\Ask-\Eff= (1-V)  \cdot  \delta \qquad \text{and} \qquad \md=\Eff+\delta\cdot(\frac{1}{2 }-V).
\end{align}
Assumption  \ref{ass:2} is obtained by letting $U\isdef\frac{1}{2 }-V$, which  is uniformly distributed on $[-1/2,1/2]$, and 
 assuming that the above formula remains valid if the spread differs from one tick.
\begin{theorem}\label{th:debiais}
Under assumption \ref{ass:2}, define 
\begin{align}
\rho&\isdef -\frac{\partial_P\IV^2(\md(\tau,k),\tau,k)}{ \partial^2_P\IV^2(\md(\tau,k),\tau,k)\cdot \spr(\tau,k)},
\label{eq:th:debiais:rho}
\end{align}
and assume that $|\rho|\geq \frac{1}{\sqrt{ 12}}$. Then set
\begin{align}
\nu&\isdef \rho-{\rm Sgn}(\rho) \sqrt{\rho^2-\frac{1}{12}}
=\frac{{\rm Sgn}(\rho)}{12\left(|\rho|+ \sqrt{\rho^2-\frac{1}{12}}\right)},\label{eq:th:debiais:nu}
\end{align}
and define the anchor price by $\anc(\tau,k)\isdef \md(\tau,k)+\spr(\tau,k)\cdot \nu$,
with the convention $|\rho|=+\infty$ and $\nu=0$ if $\partial^2_P\IV^2(\md(\tau,k),\tau,k)=0$.

Then the anchor price $\anc(\tau,k)$ depends only on $\Bid(\tau,k)$ and $\Ask(\tau,k)$, lies in the interval $(\Bid(\tau,k),\Ask(\tau,k))$,  and its bias regarding the effective price is negligible at order 3 w.r.t. the tick size:

\begin{align}\label{eq:1}
\Esp{\IV^2(\anc(\tau,k),\tau,k)}= \IV^2(\Eff(\tau,k),\tau,k)+O(\ts^3).
\end{align}
\end{theorem}
The proof is postponed to Appendix \ref{appendix:proof:theo:bias}. 
Figure \ref{fig:boxplot:anchor} depicts the boxplots of implied volatilities based on the mid-price and on the anchor price. We observe that, as expected, the anchor price has an implied volatility that is much better aligned (in mean) with the efficient IV.  Interestingly, we also note that the fluctuations of the anchor IV is slightly smaller than those of the  mid-price IV, suggesting that the use of the anchor price yields better variance properties compared to the mid-price.

\remark{The most natural choice for the distribution of $U$ is uniform on $[-\frac{1}{2 },\frac{1}{2 }]$ conditionally to $\spr$, in view of Lemma \ref{lemma:round:off:error}. We explore what results alternative modeling choices would have given regarding the bias correction from Theorem \ref{th:debiais}. 
Had we considered another distribution for $U$,  we would have obtained 
\begin{align}
\nu&\isdef 
\frac{\sigma^2\ {\rm Sgn}(\rho)}{|\rho|+ \sqrt{\rho^2-\sigma^2}}\quad\text{where}\quad\sigma\isdef\sqrt{\mathbb{V}{\rm ar} (U)},\label{eq:th:debiais:nu:alter}
\end{align}
with the constraint $|\rho|\geq \sigma$. If $\sigma>1/\sqrt {12}$, then the constraint on $\rho$ is stronger than the one in Theorem \ref{th:debiais}. 
The extremal $\nu$ (corresponding to largest unbiasing parameter) is obtained for $\rho=\pm\sigma$, for which $\nu=\pm\sigma$. Besides, the largest $\sigma$ corresponds to the Bernoulli distribution ($\sigma=1/2$).}

In Appendix \ref{subsection:verif:condition:rho}, we study under what conditions the constraint $|\rho|\geq 1/\sqrt{12}$ from  Theorem \ref{th:debiais} holds. Although at first sight this constraint 
seems quite restrictive w.r.t. $k$, it is not so if we take care to restrict the considered market data to prices greater than 1 tick, which anyone would do within a calibration framework.

\subsection{Data augmentation to handle missing data}
\label{subsection:Data augmentation when missing data}
\paragraph{About missing data.}
As previously observed (see Section \ref{section:missing_bid}), on 10\% to 15\% of the data, the bid price is outside the arbitrage bounds. With the filter on the OTM options, this corresponds to very low prices at 1 or even zero ticks  (when the data is missing). In this case of fully incoherent or missing data, a first approach is to project the observed bid price on the lower arbitrage bound rounded up to the tick. Once this is done, it is then possible to compute a mid-price that can be used as an input for the calibration process. 
However, with a large tick size and/or large spreads, the mid-price is a quite noisy representative of the efficient price and we propose a method that better takes into account all the uncertainty around the efficient price.

Generally speaking, the strategy of replacing missing or incoherent data by a reasonable guess (here, replacing the wrong bid by the arbitrage bound) is one of the possible methodologies that can be used when such an event occurs. There is an abundant literature on this topic in other fields (clinical trials, sample surveys, agricultural experiments,\dots), see for instance \cite{ende:10} for a broad overview, and to the best of our knowledge, there is no such related work in the context of volatility calibration.
It is known that replacing missing volatility data by a single value is simple but not without risk. Among the possible methods, one can take advantage of past implied volatility data, in order to naively replace the missing or incoherent bid (over the period considered for the calibration) by  a recent value. This method, which is called \emph{Last Observation Carried Forward}, does not fit well with the calibration framework because of the speed at which data changes (in finance, and especially on crypto-markets) and the use of past data does not give the best up-to-date estimation on the current implied volatility. For instance, it could be the case that in the recent past the spread was small but, due to new market conditions, the spread gets much higher with no available bid at the next calibration step, which would result in a likely large error if the recent bid was used.

Other methods using parametric predictions based on past data may suffer from the same issue of quickly changing data. Alternatively, in the 70's Rubin developed a statistical methodology based on multiple imputations, i.e. each missing value is replaced by two or more imputed values in order to represent the uncertainty about which value to impute. See \cite{rubi:87} for a monograph by this  author, and \cite{tsia:07} for a more recent reference where an asymptotic analysis is made in parametric model. This strategy allows for better robustness when missing data and improved variance due to effect of data augmentation.

\paragraph{Data augmentation.} Our approach follows the same idea of multiple imputation, i.e. data augmentation, but we use this principle not to represent the uncertainty around the bid but directly the uncertainty around the efficient price. To allow for a general treatment,  we denote by $\pie_k$ the distribution of augmented data at the log-forward-moneyness $k$; this distribution has a support on the interval\footnote{Here, $\Bid(\tau,k_i)$ is the observed bid, or the lower arbitrage bound rounded up to the tick in case of a missing or incoherent bid} $(\Bid(\tau,k_i), \Ask(\tau,k_i))$ or equivalently $(0,1)$ after a change of variable.  Thus, the initial calibration problem \eqref{eq:calibration:pb} becomes
\begin{align}\label{eq:calibration:pb:new}
\inf_\chi \frac{1}{ n} \sum_{i=1}^n \int_0^1 \Big(\tau \cdot \IV^2(\Bid(\tau,k_i)+ x\cdot \spr(\tau,k_i) ,\tau,k_i)-\omega(\tau,k_i;\chi)\Big)^2 \pie_{k_i}(\dd x).
\end{align}
For a given $k$, if we were using only the mid-price, then $\pie_k$ would be a Dirac measure at $1/2$; if we were equally using the bid and the ask prices, then $\pie_k$ would be the average of two Dirac measures (one at 0, the other at 1). One could take $\pie_k$ as the uniform distribution on $(0,1)$, but this distribution would have the same mean as the mid-price and thus, in view of Theorem \ref{th:biais}, it would yield the same bias, i.e. a strong underestimation of the effective IV in the wings (see Figure \ref{fig:bias:viz}).

\paragraph{Data augmentation with the anchor price.}
To account for the bias, we impose the distribution $\pie_k$ to be centered on the anchor price $\anc(\tau,k)$, so that
\begin{align}\label{eq:calibration:enriched:anchor}
\int_0^1 \IV^2(\Bid(\tau,k_i)+ x\cdot \spr(\tau,k_i) ,\tau,k_i) \pie_{k_i}(\dd x) =
\IV^2(\anc(\tau,k_i) ,\tau,k_i), 
\end{align}
and the final calibration is expected to have a reduced bias.

 All in all, the method that we propose, called the \emph{Data Augmentation} method, combines two effects: multiple imputations of the possible price values in the bid-ask interval for a better robustness when there is uncertain and missing  data, 
 and alignment in mean to the anchor price to reduce the calibration bias.
Although this method is motivated by missing or incoherent bids, we recommend using this method in all cases, as confirmed by our numerical tests in Section \ref{section:Numerical tests}. 

For practical considerations, there remains to define a distribution with support in $(0,1)$ with a prescribed statistics like \eqref{eq:calibration:enriched:anchor}. We propose the use of a Beta distribution ${\rm Beta}(a,b)$, the density of which is
\begin{align}
f_{a,b}(x):=\frac{x^{a-1}(1-x)^{b -1}}{B(a,b) }\1_{0<x<1} \quad \text{with}\quad B(a,b):=\frac{\Gamma(a)\Gamma(b)}{\Gamma(a+b) },
\end{align}
with some parameters $a>0,b>0$; it has  the advantage of including the uniform distribution ($a=b=1$), and being easily tunable because it is defined by two parameters only. Since the target in \eqref{eq:calibration:enriched:anchor} depends on the anchor price, the calibrated parameters $a,b $ will depend on the log-forward-moneyness $k_i$, and on the bid and ask prices.
Finding parameters satisfying  \eqref{eq:calibration:enriched:anchor} is made possible thanks to the following result (proved in Appendix \ref{appendice:proof:prop:loi:beta}).
\begin{proposition} \label{prop:loi:beta}
Denote by $\beta_{a,b}$ the random variable having the  ${\rm Beta}(a,b)$ distribution. For any strictly increasing continuous function $\varphi:[0,1]\mapsto \R$, the mapping
\begin{align}
a\in(0,+\infty) \mapsto \bar \varphi(a):=\Esp{\varphi(\beta_{a,1})}
\end{align}
is continuous, strictly increasing from $\varphi(0)$ to $\varphi(1)$.
\end{proposition}
As a consequence, by setting $\varphi(x)\isdef \IV^2(\Bid(\tau,k_i)+ x \cdot \spr(\tau,k_i),\, k_i,\,\tau)$ (which is strictly increasing in $x$), there is a unique $a_i\in (0,+\infty)$ such that \eqref{eq:calibration:enriched:anchor} holds: indeed the range of $\varphi(\cdot)$ is the interval $[\IV^2(\Bid(\tau,k_i),k_i,\tau), \IV^2(\Ask(\tau,k_i),k_i,\tau)]$ and the value 
$\IV^2(\anc(\tau,k_i),k_i,\tau)$ lies strictly in this interval. 
Because 
$$ \Esp{\IV^2(\Bid(\tau,k_i)+\beta_{a,1} \cdot \spr(\tau,k_i) ,k_i,\tau)}=\int_0^1 \IV^2(\Bid(\tau,k_i)+ x\cdot \spr(\tau,k_i) ,\tau,k_i) \pie_{k_i}(\dd x)$$ is strictly increasing in $a$, one can use a bisection method to find the (unique) root of \eqref{eq:calibration:enriched:anchor}.

\begin{figure}[t]
    \begin{center}
        \includegraphics[scale=0.6]{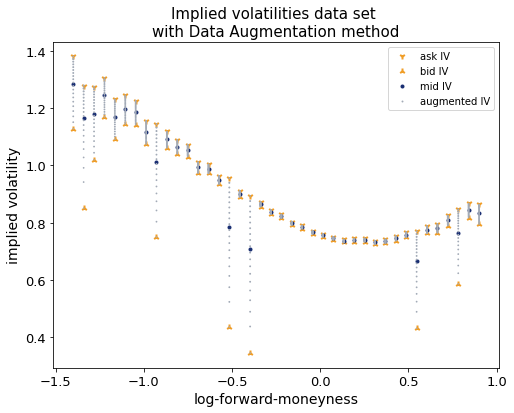}
        \caption{Illustration of the data augmentation method}
        \label{fig:sim:data_aug}
    \end{center}
\end{figure}

The principle of the data augmentation step is illustrated in Figure \ref{fig:sim:data_aug}, in the case where there are large spreads and corrupted bids. Instead of considering a continuous distribution of points in the bid-ask interval at each log-forward-moneyness, we impute a finite number of points (here 20), with calibration weights associated to the Beta distribution, tuned as above.

\section{Numerical tests}
\label{section:Numerical tests}
\subsection{Results on synthetic data}

\paragraph{SVI parametrization.}
SVI is a quite popular parametrization of implied variance which is key in describing  call and put option prices at a given expiry $\tau$.  This is a five-parameters model, $\chi^R=(a,b,\rho, m,\sigma)$, on the implied total variance, this writes
\begin{equation}
\label{eq:raw:par} \rw:=a+b\{\rho(k-m)+\sqrt{(k-m)^2+\sigma^2}\}.
\end{equation}
Here $k=\log(F_{\tau}/K)$ is the log-forward-moneyness, which depends on the forward with expiry $\tau$.

The SVI parametrization was first introduced in \cite{gath:04} at Merrill Lynch in 1999. The consistency of the SVI parametrization with large log-moneyness interpolations is studied in \cite{jackel:hyp}. In \cite{gath:jacq:11}, it is shown that the implied Heston volatility curve in large maturities is of  SVI-type (hence the name SVI for \emph{Stochastic Volatility Inspired}). Issues related to numerical calibration of an SVI parameterization are studied in \cite{dema:mart:09}. Works about interpolation  without arbitrage are available in \cite{fengler:arbitrage,glaser:arbitrage}.  There are 3 known SVI parameterizations (Raw, Natural, Jump-Wing) that we can switch between and they are associated to different parameter interpretations; in \eqref{eq:raw:par}, we choose the Raw SVI parameterization. See  \cite{gatheral:volsurface:11} for a broad overview of quantitative analysis on volatility.

In the SVI calibration, we must have $a\in \R$, $b\geq 0$, $|\rho|<1$, $m\in \R$, $\sigma>0$ and the condition $a+b\sigma\sqrt{1-\rho^2}\geq 0$ ensures that $\rw\geq 0,\ \forall k\in \R$. 
In addition, a certain function (depending on the calibrated parameters) has to be positive to ensure no butterfly arbitrage, see \cite[Section 2.2]{gath:jacq:14} for details.

For the upcoming tests on synthetic data, we use the SVI IV of Figure \ref{fig:raw:svi}.

\begin{table}[t]
    \centering
    \begin{tabular}{cccc}
        \toprule
        Parameters & \# of log-forward-moneyness & spread level & \% of spurious bids\\
        \otoprule
       \emph{Use case 1} & 30 & 2 ticks & 0\%\\
       \emph{Use case 2} & 20 & 4 ticks & 15\%\\
       \emph{Use case 3} & 10 & 10 ticks & 30\%\\
        \bottomrule
    \end{tabular}
    \caption{Parameters used for each use case.}
    \label{tab:use_case:parameters}
\end{table}

\paragraph{Results on missing data.} To show the efficiency of the \emph{Data Augmentation} method, we compared it to the \emph{Mid} method on  three different use cases: 
\begin{enumerate}
    \item The first one, hereafter referred to as \emph{Use Case 1}, represents a standard market case with a high level of liquidity. 
    \item The second use case (\emph{Use Case 2}), represents a market with poor liquidity.
    \item The last use case (\emph{Use Case 3}) finally represents a chaotic market and a liquidity crisis.
\end{enumerate}
For each use case, we initialize test parameters to describe the market they should represent. These parameters are:
\begin{itemize}
    \item The number of log-forward-moneynesses simulated, i.e. the number of available strikes.
    \item The bid/ask spread level.
    \item The percentage of spurious bid prices (representing the bids at 0 or 1 tick observed in Section \ref{section:missing_bid}). 
\end{itemize}
Note that the last parameter acts more as an aggravating factor for the liquidity. The missing or abnormal bid prices are not removed but projected onto the arbitrage boundary (rounded up to the next tick).
Table \ref{tab:use_case:parameters} provides the values of the parameters for each use case.
\begin{table}[t]
    \centering
    \begin{tabular}{cccc}
        \toprule
        Use Case & Statistics & L1 error & L2 error\\
        \otoprule
        \multirow{2}{*}{\emph{Use Case 1}} &Mid method & 17.21 ($\pm$ 0.60) & 26.68 ($\pm$ 1.04) \\
        &Data Augmentation method & 13.06 ($\pm$ 0.41) & 19.62 ($\pm$ 0.71)\\
        \otoprule
        \multirow{2}{*}{\emph{Use Case 2}}&Mid method & 17.21 ($\pm$ 0.60) & 26.68 ($\pm$ 1.04) \\
        &Data Augmentation method & 13.06 ($\pm$ 0.41) & 19.62 ($\pm$ 0.71)\\
        \otoprule
        \multirow{2}{*}{\emph{Use Case 3}}&Mid method & 69.14 ($\pm$ 2.76) & 95.78 ($\pm$ 3.88) \\
        &Data Augmentation method & 49.41 ($\pm$ 1.55) & 67.69 ($\pm$ 2.34)\\
        \bottomrule
    \end{tabular}
    \caption{Mean of the L1 and L2 errors and their 95\%-confidence intervals}
    \label{tab:res_case123}
\end{table}
Apart from the first step of simulating the 1000 samples, the rest of the test is identical for each case:
\begin{itemize}
    \item Computing the mid-prices and their associated implied volatilities.
    \item Augmenting the data in the bid-ask interval by 100 equally spaced points and weighted with the  Beta distribution, see Section \ref{subsection:Augmenting data in the bid-ask interval}, for the \emph{Data Augmentation} method. Note that we have also tested adding only 10 data points and we do not observe any  significant difference. 
    \item Calibrating two SVI smiles using BFGS for the optimization algorithm \cite[Chapter 3]{flet:optim:1} and the Huber loss function \cite{hube:64}, on both data sets (the one obtained using the \emph{Mid} method and the other using the \emph{Data Augmentation} method).
    \item Computing calibration errors over all log-forward-moneynesses between the calibrated implied volatilities and the \emph{efficient} ones for each method:
    \begin{itemize}
        \item The bias: mean of the differences between the volatilities.
        \item The L1 error: mean of the absolute values of the differences between the volatilities.
        \item The L2 error: quadratic mean of the differences between the volatilities. 
    \end{itemize}
\end{itemize}

\begin{figure}[th]
    \begin{center}
        \includegraphics[scale=0.35]{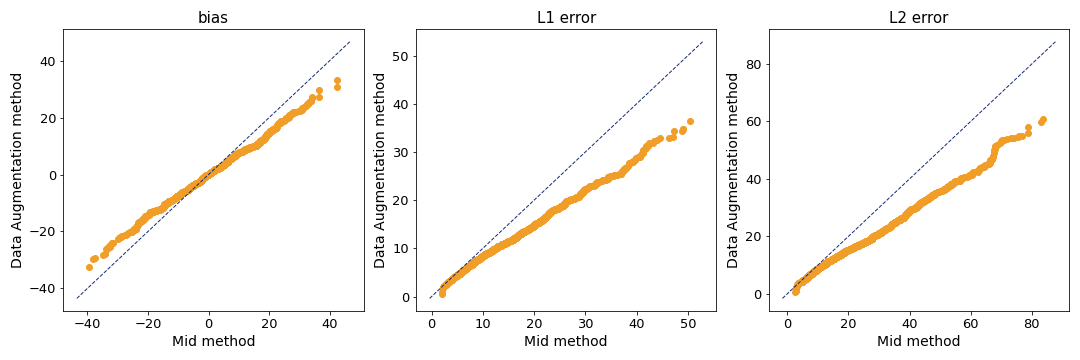}
        \includegraphics[scale=0.35]{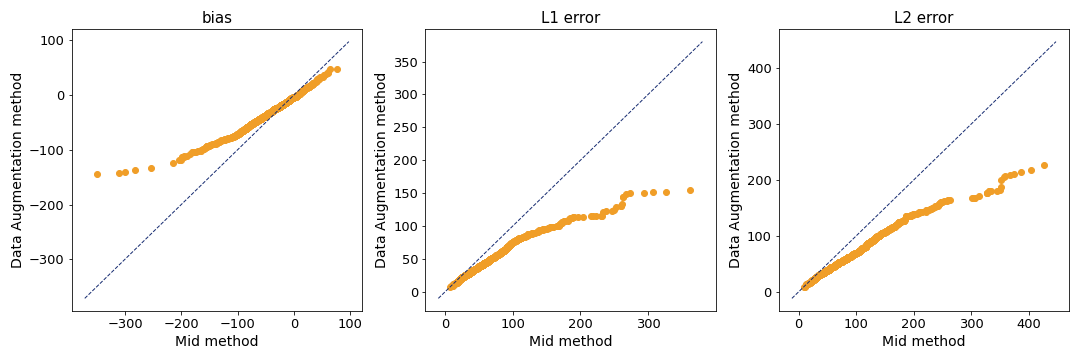}
        \includegraphics[scale=0.35]{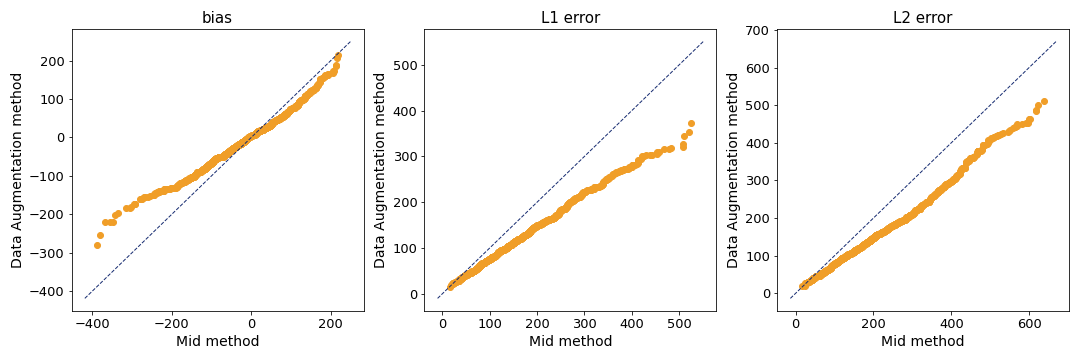}
        \caption{Q-Q plot of the calibration errors for all samples. Top row: \emph{Use Case 1}, middle row: \emph{Use Case 2}, bottom row: \emph{Use Case 3}.}
        \label{fig:qqplot:usecase123}
    \end{center}
\end{figure}

\begin{figure}[th]
    \begin{center}
        \includegraphics[scale=0.35]{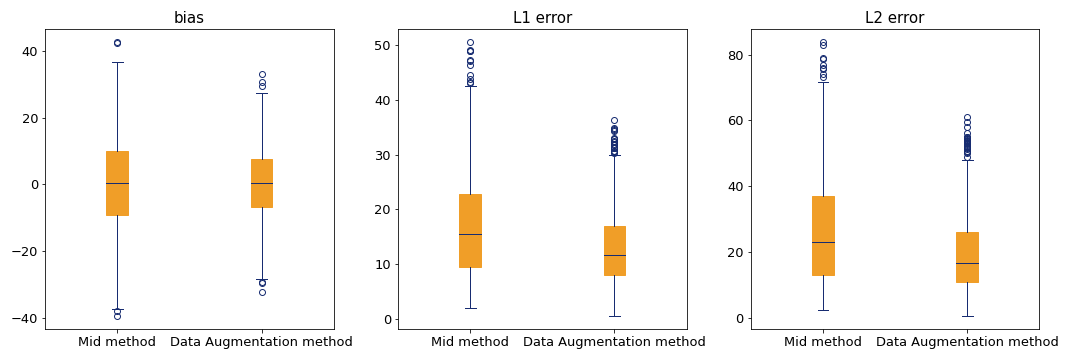}
        \includegraphics[scale=0.35]{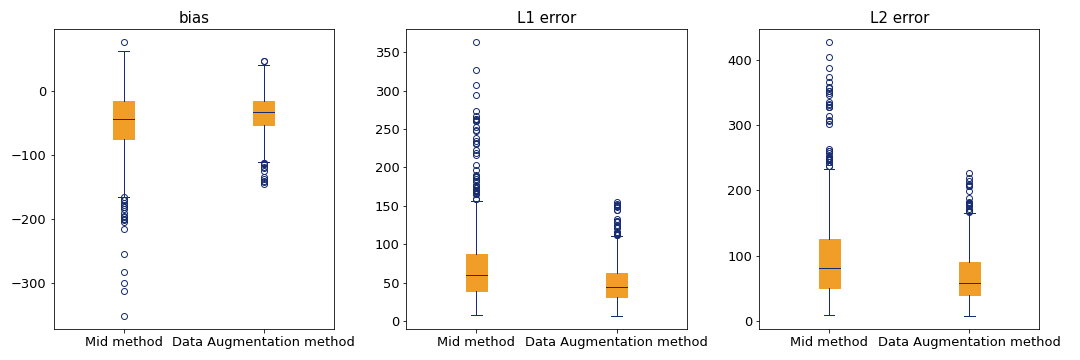}
        \includegraphics[scale=0.35]{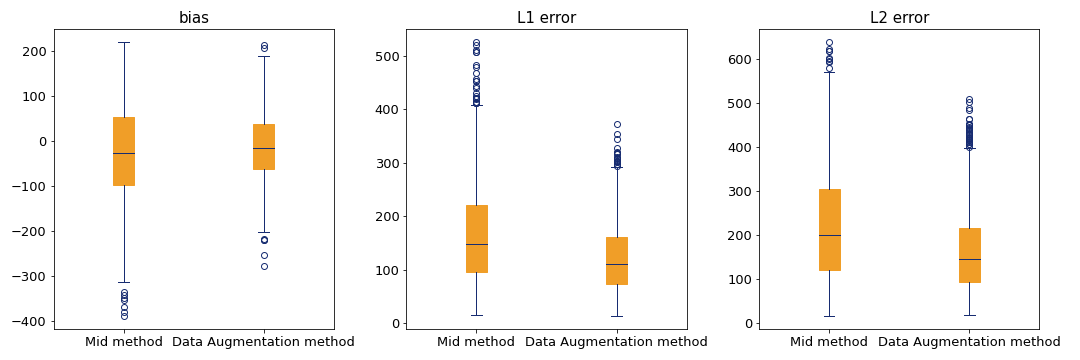}
        \caption{Box plots of the L1 and L2 errors for all samples. Top row: \emph{Use Case 1}, middle row: \emph{Use Case 2}, bottom row: \emph{Use Case 3}.}
        \label{fig:boxplot:usecase123}
    \end{center}
\end{figure}

The following tables and figures show the calibration errors and some statistics. All results are expressed in bps. 
\begin{itemize}
    \item Table \ref{tab:res_case123} contains the  error averages over the 1000 scenarii, along with  their confidence intervals.
    \item In Figure \ref{fig:qqplot:usecase123}, we plot the Q-Q plots of the bias and the L1 and L2 errors over the 1000 scenarii.
    \item Figure \ref{fig:boxplot:usecase123} presents the Box plots of the L1 and L2 errors corresponding to the 1000 scenarii.
\end{itemize}

Table \ref{tab:res_case123}  shows that the errors are significantly reduced. The L1 error is down by 24\% (resp. 29\% and 27\%) for \emph{Use Case 1} (resp. \emph{Use Case 2} and \emph{Use Case 3}) and the L2 error is reduced by 26\% (resp. 29\% and 30\%). This reduction by on average 27.5\% (regardless of the use case, i.e., the number of log-forward-moneynesses, the spread or the percentage of missing bids) shows the efficiency of the \emph{Data Augmentation} method and also confirms its stability. We have conducted other tests with different use-cases, results on the performance are not reported here but they are similar.

In the Q-Q plots we note that the L1 and L2 errors are much lower with the \emph{Data Augmentation} method than with the \emph{Mid} method, and this applies to the mean (already reported in Table \ref{tab:res_case123}) but also to the whole distribution which is closer to zero. As for the bias, whether positive or negative, it is more concentrated around zero with the \emph{Data Augmentation} method.

Finally the Box plots show that the problematic cases (those with the largest errors) are more limited with the \emph{Data Augmentation} method, since the  dispersion of the errors is smaller and the extreme values are lower. This shows the better robustness of this method over the \emph{Mid} method. 

\subsection{Results on real data}
\begin{figure}[t]
    \centering
    \includegraphics[scale=0.4]{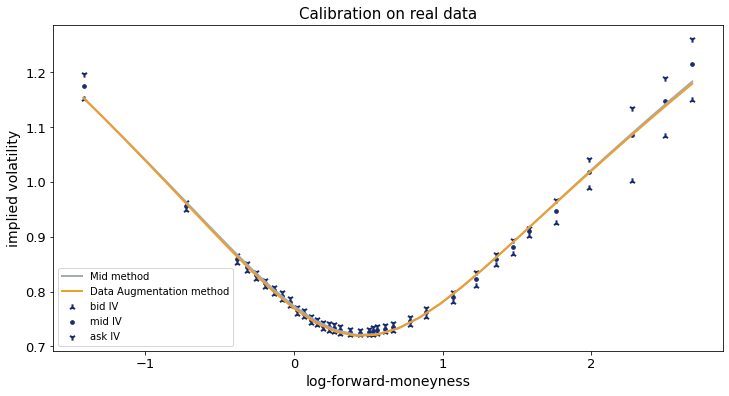}
    \caption{Calibration of quotes for BTC-USD options with 2022/12/30 expiry. Quotes are extracted from the order book snapshots of June 22, 2022 at 17:07:00 CEST on the Deribit platform.}
    \label{fig:calib:std_case}
\end{figure}

The main goal of this work is the design of an implied volatility calibrator for cryptocurrency options that does not require any human intervention to generate consistent, robust and stable smiles over time.
In most cases, enriching trades with bid-ask quotes and running the calibration process  with trades and mid-prices is sufficient to construct reasonable smiles; for these situations such as the one in Figure \ref{fig:calib:std_case}, it is not clear how to quantify the gain that is obtained  using the  \emph{Data Augmentation} method. This method appears qualitatively similar on this example but the theoretical analysis allows us to expect an overall benefit on all smiles that are calibrated.

\begin{figure}[t]
    \centering
    \includegraphics[scale=0.4]{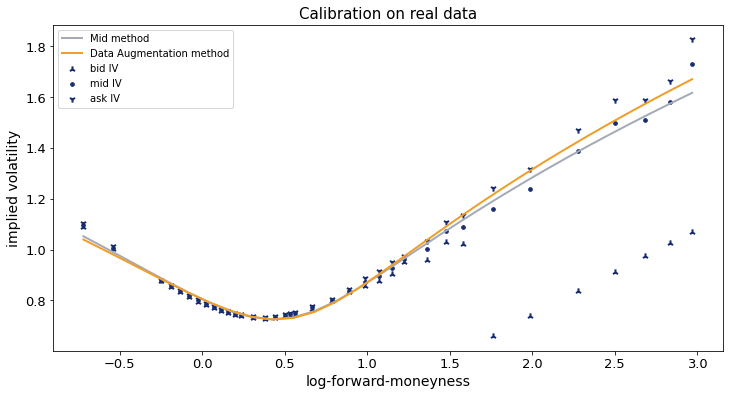}
    \caption{{Calibration of quotes for BTC-USD options with 2022/09/30 expiry. Quotes are extracted from the order book snapshots of June 22, 2022 at 17:07:00 CEST on the Deribit platform.}}
    \label{fig:calib:case_missing_data}
\end{figure}

In other cases, such as those involving  liquidity issues (large bid-ask spreads and missing or incoherent bid prices), the \emph{Data Augmentation} method corrects the bias due to these liquidity problems. Figure \ref{fig:calib:case_missing_data} depicts one of these cases. One can note that for all log-forward-moneynesses above 1.6, bid prices are missing and are replaced by the minimum price level, i.e. 1 tick. As a consequence, mid-prices are skewed downward and implied volatilities are underestimated. The \emph{Data Augmentation} method allows to correct this. 

\section{Conclusion}
Cryptocurrency options are growing increasingly popular, both for retail and institutional investors. This has led to a higher demand for tools that are standard in traditional markets, but less so in cryptocurrency markets, such as calibrators for implied volatilities. Although derivative products --and in particular options-- that can be traded on cryptocurrency markets are similar to those that are available on traditional ones, standard calibration procedures cannot be applied in a straightforward way to crypto options. Indeed, there are major differences between both markets, such as relative tick sizes, bid-ask spread sizes and amount of missing data that entail a problematic loss of accuracy in the calibrated implied volatilities.

In this paper we have designed a new calibration procedure specifically designed to overcome the shortcomings of standard procedures based on trade or mid-prices. Our calibration procedure is based on two features:
\begin{itemize}
\item the construction of an anchor price that permits to get rid of the bias that occurs in the wings for calibration procedures based on mid-prices;
\item a data augmentation step that permits to handle in a robust way the case where some of the prices to be used in the calibration are missing or incoherent.
\end{itemize}
The experimental results from Section \ref{section:Numerical tests} show that the combination of both features results in a calibration procedure that provides more accurate results than standard ones.

A natural direction for future work will consist in investigating how our procedure can be extended to handle the calibration of implied volatility surfaces.

\paragraph{Acknowledgments.}
The authors wish to thank Clara Medalie and Conor Ryder from Kaiko for their valuable feedback, as well as the Kaiko teams for their inputs and the data they provided. 

\appendix
\section{Technical results}
\label{section:appendice}

\subsection{Sensitivity of $\IV$}
\begin{lemma}\label{lemma:IV:derivative} Let $\tau,k$ be fixed. Then the implied volatility function $P\mapsto \IV(P)$ as a function of the input price  $P$ is such
\begin{align}\label{lemma:IV:derivative:1}
\partial_P (\IV^2)&=2\,\IV /\vega,\\
\label{lemma:IV:derivative:2}
\partial^2_P(\IV^2)&=\frac{2}{\vega^2 }\left( -\frac{k^2}{\IV^2\tau }+\frac{\IV^2\tau}{4 } +1\right),
\end{align}
where $\vega$ is the Black-Scholes Vega with parameters $\tau,k, \IV$.
\end{lemma}
\begin{proof}
Writing $P={\tt Black\text{-}Scholes~price}(\IV(P))$ and omitting the indices $\tau,k,\BS$ for the sake of simplicity, we have $1=\vega\cdot \partial_P \IV(P)$ and $0=\vomma\cdot (\partial_P \IV(P))^2+ \vega\cdot \partial^2_P \IV(P)$. Using $\vomma=\vega \frac{d_+ d_-}{ \IV}$ \cite[p.23]{haug:07} with the usual definitions for $d_\pm$, we get
\begin{align*}
\partial_P \IV(P)&=1/\vega,\qquad \partial^2_P \IV(P)=-\vomma/\vega^3=-d_+ d_-/(\IV\cdot\vega^2 ),\\
\partial^2_P(\IV^2)&=\frac{2}{\vega^2 }( -d_+ d_- +1)=\frac{2}{\vega^2 }\left( -\frac{k^2}{\IV^2\tau }+\frac{\IV^2\tau}{4 } +1\right).
\end{align*}
\end{proof}

\subsection{Distribution of round-off error}
The following result states a nice and explicit approximation of the joint distribution of a random variable and its round-off error, in the asymptotics of small round-off level. Such results were initiated by \cite{kosu:37} and \cite{tuke:39}, and stated in a refined form later  in \cite{dela:jaco:97}. Define the fractional part of a scalar $x$ as follows:
\begin{align}\label{eq:fractional:part}
\{x\}=x-\sup\{n\in \N:n\leq x\}.
\end{align}
\begin{lemma}\label{lemma:round:off:error}
Let $\delta$ be a round-off level and let $Y$ be a scalar random variable, which distribution is assumed to have a density which is $N$-times differentiable $(N\geq 1$) with integrable derivatives.  Then, for any function $f$ assumed to be $N$ times differentiable with bounded derivatives, we have 
\begin{align}
\label{eq:round:off:error}
\Esp{f(Y,\{Y/\delta\})}-\Esp{f(Y,V)} ={\cal O}(\delta^N),
\end{align}
where $V$ is a random variable, uniformly distributed on $[0,1]$ and independent from $Y$.
\end{lemma}
In words, the round-error of $Y$ defined by $\delta\{Y/\delta\}$ is asymptotically independent from $Y$ and uniformly distributed on $[0,\delta]$.

\subsection{Proof of Theorem \ref{th:debiais}}
\label{appendix:proof:theo:bias}
We omit the index $(\tau,k)$ in the notations. 
A second-order Taylor expansion of  $\IV^2(.)$ gives, keeping in mind that $|\Eff-\md|\leq \frac{1}{2 }\spr$, $|\anc-\md|\leq \frac{1}{2 }\spr$ and $\spr=O(\ts)$:
\begin{align*}
{\IV^2(\anc)}&={\IV^2(\Eff)+\partial_P\IV^2(\Eff)(\anc-\Eff)+\frac{1}{2 }\partial^2_P\IV^2(\Eff)(\anc-\Eff)^2}+O(\ts^3).
\end{align*}
Because $\anc-\Eff=\spr\cdot (\nu+ U)$, observe that \eqref{eq:1} is equivalent to 
\begin{align}
0&=\Esp{\partial_P\IV^2(\Eff)\cdot\spr\cdot (\nu+ U)+\frac{1}{2 }\partial^2_P\IV^2(\Eff)\cdot \spr^2 \cdot(\nu+U)^2}+O(\ts^3)\\
&=\Esp{\partial_P\IV^2(\Eff)\cdot\spr\cdot \nu+\frac{1}{2 }\partial^2_P\IV^2(\Eff)\cdot\spr^2 \cdot(\nu^2{+ 2\nu \cdot U}+\frac{1}{12 })}+O(\ts^3),
\label{eq:3}
\end{align}
where we used the fact that $U$ is uniformly distributed on $[-1/2,1/2]$, conditionally to $\spr$.
{Note that this conditioning and centering argument does not apply to the term $2\nu\cdot U$, since $\nu$ depends on $(\spr,\md)$ 
and not on $\spr$ only. However, observe that  
\begin{align}
\Esp{
\partial^2_P\IV^2(\Eff)\cdot\spr^2 \cdot U \cdot \nu(\md,\spr)}=O(\ts^3).
\label{eq:3:inter}
\end{align}
 Indeed the function $\nu$ is smooth in the variable $\md$, so that $\nu(\md,\spr)=\nu(\Eff,\spr)+O(\ts)$, and since $U$ is centered conditionally to $\spr$ we get 
$$\Esp{
\partial^2_P\IV^2(\Eff)\cdot\spr^2 \cdot U \cdot \nu(\md,\spr)}=\Esp{
\partial^2_P\IV^2(\Eff)\cdot\spr^2 \cdot U \cdot \nu(\Eff,\spr)}+O(\ts^3)=O(\ts^3).$$
 
It follows that
\begin{align}
0&=\Esp{\partial_P\IV^2(\Eff)\cdot\spr\cdot \nu+\frac{1}{2 }\partial^2_P\IV^2(\Eff)\cdot\spr^2 \cdot(\nu^2+\frac{1}{12 })}+O(\ts^3).
\label{eq:3:bis}
\end{align}
}
Moreover, writing \begin{align}
\partial^2_P\IV^2(\Eff)&=\partial^2_P\IV^2(\md)+O(\ts),\\
\partial_P\IV^2(\Eff)&=\partial_P\IV^2(\md)+\partial^2_P\IV^2(\Eff)(\Eff-\md)+O(\ts^2),
\end{align}
\eqref{eq:3:bis} becomes
\begin{align}\label{eq:4}
0&=
\Esp{\partial_P\IV^2(\md)\cdot\spr\cdot \nu-
\partial^2_P\IV^2(\Eff)\cdot\spr^2 \cdot U \cdot \nu+
\frac{1}{2 }\partial^2_P\IV^2(\md)\cdot\spr^2\cdot (\nu^2+ \frac{1}{12 })}+O(\ts^3)\\
&=\Esp{\partial_P\IV^2(\md)\cdot\spr\cdot \nu+
\frac{1}{2 }\partial^2_P\IV^2(\md)\cdot\spr^2\cdot (\nu^2+ \frac{1}{12 })}+O(\ts^3),
\end{align}
using  \eqref{eq:3:inter} once again.

Hence, there remains to check that $\nu$ in \eqref{eq:th:debiais:nu} solves 
\begin{align}
0&=
\partial_P\IV^2(\md)\cdot\spr\cdot \nu+
\frac{1}{2 }\partial^2_P\IV^2(\md)\cdot\spr^2\cdot (\nu^2+ \frac{1}{12 }).
\end{align}
If $\partial^2_P\IV^2(\Eff)=0$, then $\nu=0$ solves the above and this coincides with \eqref{eq:th:debiais:nu} by taking $|\rho|\uparrow +\infty$. In the other cases, this writes
$(\nu-\rho)^2=\rho^2-\frac{1}{12 }$.
This equation has one or two roots for $|\rho|\geq 1/12$ and the one in \eqref{eq:th:debiais:nu}  corresponds to the closest to zero. 

Last we easily check that from the right hand side of \eqref{eq:th:debiais:nu}, that $|\nu|\leq 1/\sqrt 12<1/2$ so that the anchor price lies strictly in the bid-ask interval. This concludes the proof.\qed

\subsection{Verification of the condition on $\rho$ in Theorem \ref{th:debiais}}
\label{subsection:verif:condition:rho}
\begin{figure}[htb]
\begin{center}
\includegraphics[scale=0.5]{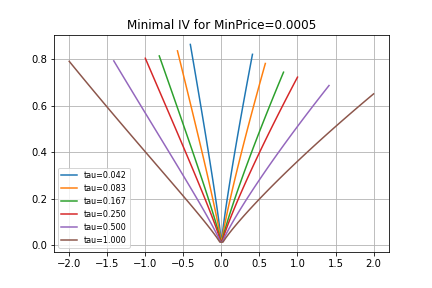}
\includegraphics[scale=0.5]{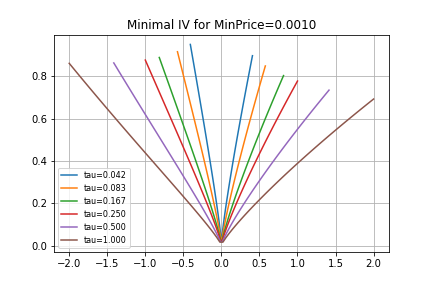}
\caption{Minimal implied volatility $\bIV$ as a function of $k$, when the minimal price is $n=1$ tick on the left and $n=2$ ticks on the right. Time-to-expiries $\tau$ are 2 Weeks, 1M, 2M, 3M, 6M, 1Y and are expressed in years.\label{fig:Minimal_IV}}
\end{center}
\end{figure}

\begin{figure}[htb]
\begin{center}
\includegraphics[scale=0.33]{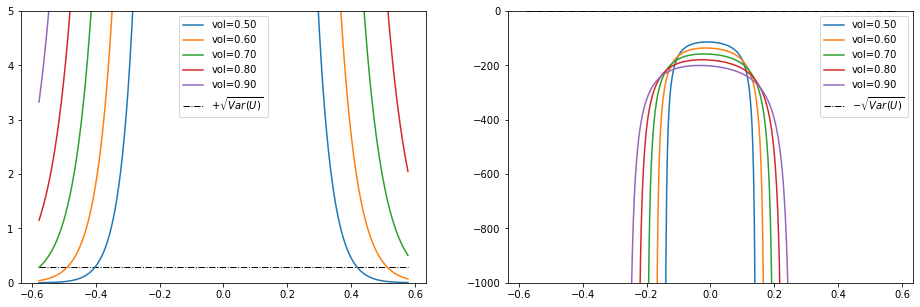}
\includegraphics[scale=0.33]{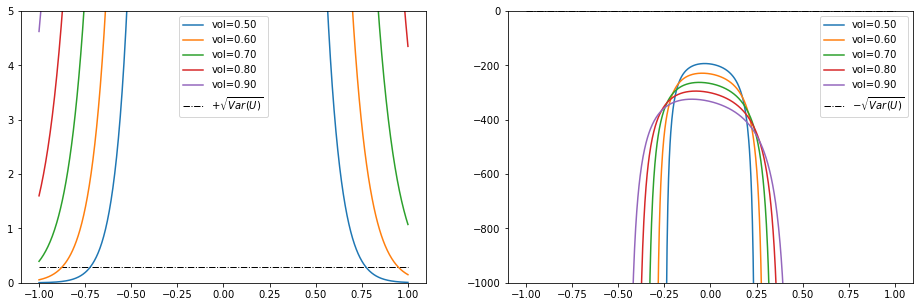}
\includegraphics[scale=0.33]{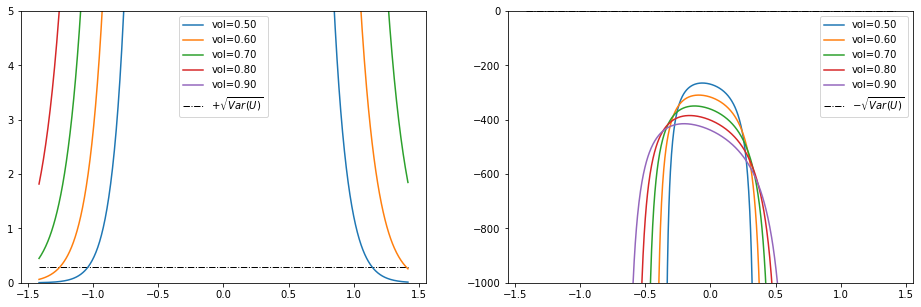}
\caption{Behavior of $\sigma\mapsto \rho(\sigma)$, depending on the log-forward-moneyness $k$, when the   spread in \eqref{eq:rho:simplified} is 1 tick. The time-to-expiry is 1 month at the top, 3 months in the middle, 6 months at the bottom. \label{fig:rho_increasing_wrt_vol}}
\end{center}
\end{figure}

\begin{figure}[htb]
\begin{center}
\includegraphics[scale=0.4]{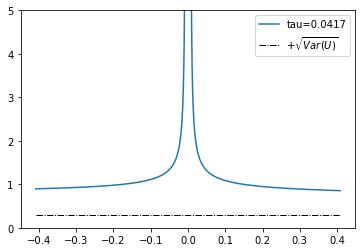}
\includegraphics[scale=0.4]{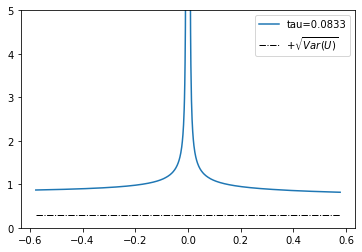}
\includegraphics[scale=0.4]{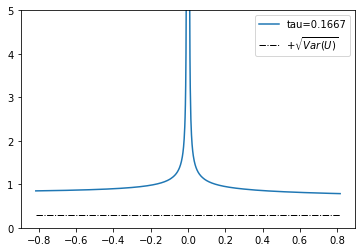}

\includegraphics[scale=0.4]{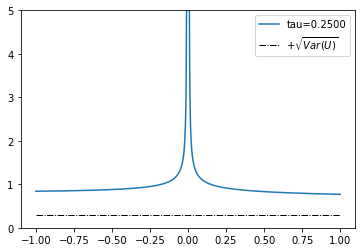}
\includegraphics[scale=0.4]{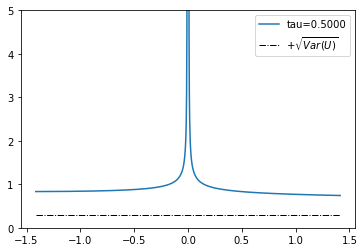}
\includegraphics[scale=0.4]{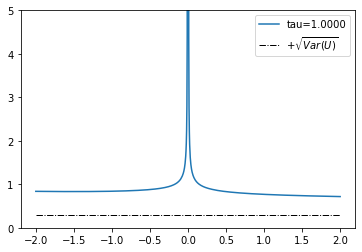}

\caption{The function $k\mapsto \rho(k)$ when the input price is computed with the most extreme volatility $\bIV(k)$ (for a minimal price of one tick). From top left to bottom right, different time-to-expiries $\tau$ in years (2W, 1M, 2M, 3M, 6M, 1Y). \label{fig:rho:bIV}}
\end{center}
\end{figure}

Let us now study under which conditions  the constraint $|\rho|\geq 1/\sqrt{12}$ from Theorem \ref{th:debiais} is met. First, we can simplify the expression of $\rho$ using Lemma \ref{lemma:IV:derivative}:
\begin{align}
\rho=-\frac{\partial_P\IV^2(\md(\tau,k),\tau,k)}{ \partial^2_P\IV^2(\md(\tau,k),\tau,k)\ \spr(\tau,k)}
&=-\frac{\IV\cdot \vega }{\left( -\frac{k^2}{\IV^2\tau }+\frac{\IV^2\tau}{4 } +1\right)\spr}.
\label{eq:rho:simplified}
\end{align}
Around the log-forward-money $k\approx 0$, the situation is quite clear: indeed, from \cite[p.23]{haug:07},  $\vega=F \sqrt \tau \cN'(d_+)$  and  neglecting $\frac{\IV^2\tau}{4 }$ for usual expiries and volatilities, we get 
{$\rho\approx -\IV\frac{F}{\spr}\sqrt{\frac{\tau}{2\pi}}\approx
-\IV\sqrt \tau \cdot 800$,}
which is much larger (in absolute value) than $1/\sqrt{12}$ for any realistic market data.
Thus, our concern is rather for OTM values of $k$. As $k\to\pm \infty$, we observe that $\rho\to0$ all other things being equal, making it seemingly hopeless to satisfy  the condition $|\rho|\geq 1/\sqrt{12}$. But in practice, the option prices that are used for calibration, are above one or two ticks;  this significantly restricts the range of possible values of $k$ and $\IV$ --at least, this imposes a constraint on $\IV$ w.r.t. $k$. More precisely, define $\bIV(k)$ as the Black-Scholes volatility for which the OTM Call or the OTM Put (with forward-log-moneyness $k$) is worth $n$ ticks. In other words, any volatility larger than $\bIV(k)$ leads to an option price larger than $n$ ticks (i.e. to a market data point that is considered for calibration) and conversely, any market data point in the calibration set (i.e. with a value above $n$ ticks) with log-forward-moneyness $k$  has an implied volatility larger than $\bIV(k)$. Figure \ref{fig:Minimal_IV} depicts this minimal Implied Volatility $\bIV$ as a function\footnote{The range of $k$ is defined as $[-2\sqrt \tau,+2\sqrt \tau]$, which easily encompasses the observed ranges of $k$ as discussed in Section \ref{section:Market data}.}  of $k$, in the case where the minimal option price is equal to $n=1$ tick (this represents the most permissive case). 
We argue that these values give a worst-case volatility for checking the condition  $|\rho|\geq 1/\sqrt{12}$. Indeed, on Figure \ref{fig:rho_increasing_wrt_vol}, we can visually check that the function 
\[\sigma \mapsto \rho(\sigma)=
-\frac{\sigma\cdot \vega(\sigma,k)}{\left( -\frac{k^2}{\sigma^2\tau }+\frac{\sigma^2\tau}{4 } +1\right)\spr}\] 
is increasing for OTM log-forward-moneyness $k$. Hence, it is sufficient for such OTM options to check the condition $|\rho|\geq 1/\sqrt{12}$ when the input price is generated by the implied volatility $\bIV(k)$. Figure \ref{fig:rho:bIV} gives the graphs of such extremal $\rho$-functions for different time-to-expiries (when the minimal price is 1 tick): in all cases, the required condition on $\rho$ is largely met. Had we only considered option with prices greater than 2 ticks or more, we would have fulfilled even more easily the condition $|\rho|\geq 1/\sqrt{12}$. 

\subsection{Proof of Proposition \ref{prop:loi:beta}}
\label{appendice:proof:prop:loi:beta}
The mean and variance of Beta$(a,b)$ distribution is 
$$\Esp{\varepsilon_i}=\frac{a}{a+b },\qquad 
\var(\varepsilon_i)=\frac{ab}{(a+b)^2 (a+b+1)}.$$
Therefore, for $b=1$, 
\begin{itemize} \item as $a\downarrow 0$, $\Esp{\varepsilon_i}\to 0$ and $\var(\varepsilon_i)\to0$, hence $\varepsilon_i\to 0$ in probability, thus $\lim_{a\downarrow 0} \Esp{\varphi(\beta_{a,1})}=\varphi(0)$.
\item as $a\uparrow +\infty$, $\Esp{\varepsilon_i}\to 1$ and $\var(\varepsilon_i)\to0$,  $\varepsilon_i\to 1$ in probability and $\lim_{a\uparrow +\infty} \Esp{\varphi(\beta_{a,1})}=\varphi(1)$. 
  \end{itemize}
 It remains to show that $\bar \varphi(a)$ is strictly increasing and continuous in $a$. The continuity is clear writing $\bar \varphi$ as an integral with respect to the density $f_{a,1}$. The increasingness can be shown using the Beta-Gamma algebra property \cite{dufr:98}:
$\varepsilon^{(a,b)}\overset{d}=\frac{\Gamma^{(a)}}{\Gamma^{(a)}+\Gamma^{(b)} }
< \frac{\Gamma^{(a)}+\Gamma^{(\varepsilon)}}{\Gamma^{(a)}+\Gamma^{(\varepsilon)}+\Gamma^{(b)} }
\overset{d}=
\frac{\Gamma^{(a+\varepsilon)}}{\Gamma^{(a+\varepsilon)}+\Gamma^{(b)} }\overset{d}=\varepsilon^{(a+\varepsilon,b)}
$
where all Gamma distributed random variables are independent; then  $\Esp{\varphi(\beta_{a,1})}<\Esp{\varphi(\beta_{a+\varepsilon,1})}$ using the strict monotony of $\varphi$, hence the announced property.\qed

\end{document}